\documentclass[a4paper]{article}

\usepackage{lineno,hyperref}

\usepackage
	{todonotes}

\usepackage[english]{babel}
\usepackage[utf8]{inputenc}
\usepackage[T1]{fontenc}
\usepackage{graphicx}
\usepackage{algorithm}
\usepackage{algorithmic}
\usepackage{mathtools}
\usepackage{amssymb}
\usepackage{mathptmx}
\usepackage{xcolor}
\usepackage{amsmath}
\usepackage{multirow}
\usepackage[margin=2cm]{geometry}

\newtheorem{definition}{Definition}

\usepackage{lastpage}
\usepackage{fancyhdr}
\usepackage{titling}

\usepackage{doi}
\usepackage[autostyle]{csquotes}
\usepackage[style=authoryear,citestyle=authoryear-comp,maxcitenames=2,maxbibnames=99,backend=bibtex,backref=true]{biblatex}
\title{Exploring the Evolution of Node Neighborhoods \\in Dynamic Networks}

\author{Günce Keziban \textsc{Orman}\textsuperscript{1}, Vincent \textsc{Labatut}\textsuperscript{2} \& Ahmet Teoman \textsc{Naskali}\textsuperscript{1}\\
	\vspace{-0.2cm}\small1: Computer Engineering Department, Galatasaray University, 
Ortaköy, Istanbul, Turkey \\
	\small\href{mailto:korman@gsu.edu.tr}{\texttt{korman@gsu.edu.tr}}, \href{mailto:tnaskali@gsu.edu.tr}{\texttt{tnaskali@gsu.edu.tr}} \\
	\vspace{-0.2cm}\small2: Laboratoire Informatique d'Avignon (LIA) EA 4128, Université d'Avignon, France \\
    \small\href{mailto:vincent.labatut@univ-avignon.fr}{\texttt{vincent.labatut@univ-avignon.fr}}}

\usepackage{hyperref}
\hypersetup{
    pdftitle={\thetitle},
    bookmarksnumbered=true,bookmarksopen=true,
	unicode=true,colorlinks=true,linktoc=all,
	linkcolor=blue,citecolor=blue,filecolor=blue,urlcolor=blue,
	pdfstartview=FitH
}

\begin{document}

\maketitle
\begin{abstract}
Dynamic Networks are a popular way of modeling and studying the behavior of evolving systems. However, their analysis constitutes a relatively recent subfield of Network Science, and the number of available tools is consequently much smaller than for static networks. In this work, we propose a method specifically designed to take advantage of the longitudinal nature of dynamic networks. It characterizes each individual node by studying the evolution of its direct neighborhood, based on the assumption that the way this neighborhood changes reflects the role and position of the node in the whole network. For this purpose, we define the concept of \textit{neighborhood event}, which corresponds to the various transformations such groups of nodes can undergo, and describe an algorithm for detecting such events. We demonstrate the interest of our method on three real-world networks: DBLP, LastFM and Enron. We apply frequent pattern mining to extract meaningful information from temporal sequences of neighborhood events. This results in the identification of behavioral trends emerging in the whole network, as well as the individual characterization of specific nodes. We also perform a cluster analysis, which reveals that, in all three networks, one can distinguish two types of nodes exhibiting different behaviors: a very small group of active nodes, whose neighborhood undergo diverse and frequent events, and a very large group of stable nodes. 

\vspace{0.3cm}
\textbf{Keywords:} Dynamic networks, Network Evolution, Network Topology, Neighborhood Events.

\vspace{0.3cm}
\textcolor{red}{\textbf{Cite as:} G. K. Orman, V. Labatut \& A. T. Naskali. \href{http://www.sciencedirect.com/science/article/pii/S0378437117304053}{Exploring the Evolution of Node Neighborhoods in Dynamic Networks}. Physica A: Statistical Mechanics and its Applications, 482:375-391, 2017. Doi: \href{https://doi.org/10.1016/j.physa.2017.04.084}{10.1016/j.physa.2017.04.084}}
\end{abstract}


\section{Introduction}
\label{intro}
Dynamic Network Analysis is a subfield of Network Science aiming at representing and studying the behavior of systems constituted of interacting or related objects evolving through time. This domain gained attention lately, as attested by the recent publication of several surveys \parencite{Kuhn2011, Spiliopoulou2011, Blonder2012, Holme2012, Aggarwal2014}.

Authors generally distinguish two types of such dynamic systems. In the first, interactions or relationships are very short, or even punctual, and occur frequently. They are called \textit{contact sequences} \parencite{Holme2012} and \textit{streaming networks} \parencite{Aggarwal2014} in the literature (among other names). In the second type, called \textit{interval graphs} \parencite{Holme2012} and \textit{slowly evolving networks} \parencite{Aggarwal2014}, interactions or relationships have a significant duration and are not so frequent. Dynamic network modeling can be applied to both types, although it is more suited to the second. It consists in representing the system evolution through a sequence of graphs. Each one of these graphs, which we call \textit{time slices} in this article, represents the aggregation of the system changes over a given period of time. This representation allows studying the system properties with sequence-based tools such as time series analysis and sequential pattern mining.

According to Aggarwal \& Subbian, one can distinguish two types of methods to analyze dynamic networks \parencite{Aggarwal2014}. On the one hand, so-called \textit{maintenance methods} consist in performing an initial analysis on the first time slice (e.g. detect the community structure), and then updating its output for each subsequent time slice (e.g. move certain nodes from one community to another). This is typically done with methods originally developed for static networks and later adapted to dynamic ones. On the other hand, \textit{evolution methods} embrace the longitudinal nature of the data and focus on describing the changes caused by temporal evolution. They are useful to identify and understand the rules governing the evolution of the studied network, and allow the design of new models. Note that it is possible to simultaneously belong to both types of methods.
Another important feature is the granularity used to perform the analysis. First, \textit{macroscopic} methods deal with the whole network at once. They are the most widespread in the literature, e.g. network density, network diameter \parencite{Leskovec2005}. Second, \textit{mesoscopic} methods consider the network at an intermediary level, generally that of the community structure, e.g. modularity measure \parencite{Kashtan2005}, size of the communities \parencite{Backstrom:2006}. Finally, \textit{microscopic} methods focus on individual nodes and possibly their direct neighborhood, e.g. clustering coefficient \parencite{Clauset2007}.

By definition, methods specifically designed to handle a dynamic network are more likely to take advantage of the specificity of such data. Regarding the granularity, if macroscopic and mesoscopic methods are widespread in the literature, it is not the case for microscopic methods. Yet, the benefits of such approaches are numerous: by allowing the tracking of finer evolution processes, they complement macroscopic and/or mesoscopic results. They help identifying behavioral trends among the network nodes, and consequently outliers. These results can facilitate the description and understanding of processes observed at a higher level, and can be useful to define models of the studied system, especially agent-based ones.

In this work, we propose a method specifically designed to study dynamic slowly evolving 
networks, at the microscopic level. We characterize a node by the evolution of its neighborhood. More precisely, we detect specific events occurring among the groups of nodes constituting this neighborhood, between each pair of consecutive time slices, which we call \textit{neighborhood evolution events}. This method constitutes our main contribution. To the best of our knowledge, it is the first attempt at describing the dynamics of a network based on such local events. 
For each node, our method outputs a sequence of categorical features corresponding to the different types of events it experienced. These features can then be analyzed with any tool capable of processing temporal categorical data, in order to extract meaningful information regarding the network dynamics. This knowledge can noticeably be used to categorize nodes, or identify trends and outliers. For illustration purposes, and as a second contribution, we analyze three real-world networks: DBLP (scientific collaborations), LastFM (social relations through musical tastes) and Enron (email exchanges). We first apply our event identification method, before using \textit{Sequential Pattern Mining} and some complementary analysis to extract higher level information.

The rest of this article is organized as follows. In Section \ref{sec:RelatedWork}, we review in further detail the existing work the most related to our method. In Section \ref{sec:Methods}, we formally define the concept of \textit{neighborhood evolution event}, and describe the method we use to detect such events in dynamic networks. In Section \ref{sec:results}, we apply our method to three real-world networks and discuss the obtained results. Finally, in Section \ref{sec:Conclusion}, we comment on the limitations of our work, how they could be overcome and how our method could be extended.
 

\section{Related Work}
\label{sec:RelatedWork}
This section does not aim at being exhaustive, but rather at presenting the various general families of methods existing to characterize dynamic graphs. A number of reviews exist which describe them in further detail, and list them in a more exhaustive way \parencite{Holme2012,Nicosia2013,Aggarwal2014}.

As mentioned in the introduction, the most straightforward way of characterizing a dynamic network is to use measures designed for static networks, by applying them to each time slice and considering the resulting time series. A number of studies adopted this approach for a variety of measures, at different granularities. To cite a few, at the macroscopic level: number of links \parencite{Akoglu2008}, size of the giant component \parencite{Leskovec2005a}, diameter \parencite{Leskovec2005,Leskovec2005a}, principal Eigenvalue of the adjacency matrix \parencite{Akoglu2008}; at the mesoscopic level: modularity \parencite{Kashtan2005}, size of the communities \parencite{Backstrom:2006}, number of communities \parencite{Aynaud2010}, numbers of motifs (predefined small subgraphs) \parencite{Braha2009}; and at the microscopic level: degree \parencite{Braha2009, Leskovec2005, Clauset2007}, clustering coefficient \parencite{Clauset2007}.

The next step is to apply tools originally designed to process static networks at each time slice, like before, but with some additional updating mechanism allowing to smooth the results. This is particularly common in the domain of community detection, like for instance in \parencite{Aynaud2010}. A related incremental approach, consisting in performing an initial computation on the first time slice and updating it at each additional time sleice, is also applied to the processing of centrality measures, such as the \textit{incremental closeness} \parencite{Kas2013}. In contrast, more recent tools go farther in the adaptation of these static methods to the analysis of dynamic networks and better describe or detect the changes caused by temporal evolution. For instance, \textcite{Gupta2011} describe a method allowing to detect the most significant changes in the distance between nodes, assuming such changes correspond to important events in the evolution of the graph.

With their method based on the identification of \textit{role dynamics}, \textcite{Rossi2012} use several such measures at once. They consider each one of them as a feature characterizing the topological position of a node in the graph. They apply non-negative matrix factorization to identify a small number of node roles, corresponding to typical evolutions of the features for the considered period of time. Each role is characterized by its dependence on certain features, and each node is more or less related to certain roles. The prevalence of a role can change with time, which can be used to characterize the network dynamics. For instance, \textcite{Rossi2012} observe that certain roles are more represented at day or at night, i.e. they are periodic, whereas others are more stationary.

A whole family of descriptive tools is based on the generalization of the notion of \textit{path} to dynamic networks. In a static graph, a path between two nodes is a sequence of adjacent links allowing to indirectly connect them The concept of \textit{time-respecting path} \parencite{Kempe2002} states that in a dynamic graph, the links must additionally be chronologically ordered. This can be used to generalize all the path-based measures originally designed for static graphs, such as the betweenness \parencite{Tang2010d} or edge-betweenness (number of shortest paths going through some node or link of interest, respectively). A similar chronological constraint was applied to spectral centrality measures such as the Eigenvector centrality \parencite{Grindrod2011,Nicosia2013}, Katz centrality \parencite{Grindrod2011} or PageRank \parencite{Lerman2010}.

Moreover, the distance between two nodes being the length of the shortest path connecting them, the same can be said of all the distance-based measures, such as: closeness \parencite{Magnien2015} (reciprocal of the average distance to a node of interest), eccentricity (maximal distance to a node of interest), average distance \parencite{Holme2012,Nicosia2013} (over all pairs of nodes in a graph), diameter \parencite{Holme2012} (longest distance in the graph) or efficiency \parencite{Magnien2015} (average of the distance reciprocal over all paths in the graph). Alternatively, certain authors prefer to consider the total duration of paths instead of their length, leading to the notion of \textit{temporal length} (by opposition to the traditional topological length). The temporal length of the fastest path is called temporal distance, and can be substituted to the topological distance in the previously mentioned measures, e.g. the \textit{temporal closeness} described in \parencite{Pan2011}, or the \textit{characteristic temporal path length} used in \parencite{Tang2010c} (which is the average temporal distance).

The concept of time-respecting path can also be used to generalize the notion of component, as in \parencite{Nicosia2012}. The number of components and the presence of a giant one are two largely studied properties in static graphs. Time-respecting paths can additionally be used to define connectivity-related measures. For instance, \textcite{Holme2005} defines the \textit{influence set} of a node of interest, as the set of nodes it can reach through such a path in a given time. Based on this, he derives his \textit{reachability ratio}: the proportion of nodes belonging to the influence set, averaged over the whole graph. The \textit{source set} is defined symmetrically in \parencite{Riolo2001} as the set of nodes able to reach the node of interest, and allows deriving a similar measure. It is also possible to monitor the evolution of the number of time-respecting shortest paths able to reach a node of interest for a given time window \parencite{Moody2002} (a measure not unlike the dynamic betweenness).

Several tools have been proposed to characterize dynamic graphs in terms of their constituting subgraphs. \textcite{Lahiri2007} use a pattern mining approach to detect frequent subgraphs based on their temporal support, i.e. depending on how persistent these subgraphs are. Another family of works tried to generalize the notion of motif to dynamic networks. A \textit{motif} is a generic subgraph of a few nodes appearing frequently in a static network \parencite{Milo2002}, comparatively to some null model. A number of such generalizations have been proposed: some authors detect static subgraphs in time slices \parencite{Braha2009}, others detect dynamic time-respecting subgraphs \parencite{Bajardi2011}. See Section 4.10 of \parencite{Nicosia2013} for a more exhaustive review of motif-based approaches in dynamic networks.

A number of works have transposed concepts developed from the field of data mining to the analysis of graphs. Authors generally suppose that a node can be represented by a set of descriptor values, a descriptor being either topological (e.g. degree, centrality measure, etc.) or attribute-based (e.g. for a social network: age, gender, interests of the considered person). \textcite{Berlingerio2009} propose a method to identify \textit{graph evolution rules}, i.e. association rules describing the evolution of local connections. Briefly, these rules state that if a certain local interconnection pattern is observed, the network will locally evolve in a certain way. For instance, in the case of an academic co-authorship network, such a rule could be that if a high degree individual is connected to four medium degree researchers, he is likely to get connected to a fifth one in the next time slice.

Other data mining-related works have focused on pattern mining to describe dynamic networks. In the context of traditional data mining, a pattern is a set of descriptor values common to certain instances of the studied dataset, and possessing certain properties. For instance, a \textit{frequent} pattern is shared by a large number of instances. \textcite{Desmier2012} study co-evolution patterns, i.e. the trend for certain groups of nodes to behave similarly in terms of how their descriptor values change with time. A \textit{sequential} pattern aims at describing time series, in which one looks for sequences of sets of descriptor values \parencite{Mabroukeh2010}. \textcite{Orman2014a} use \textit{sequential} pattern mining to make sense of dynamic communities and ease their interpretation.

Most of the methods defined at the level of the community structure focus on so-called \textit{community events}. These were originally proposed by \textcite{Toyoda2003}, and have later been used by many other authors. They correspond to the $6$ different kinds of transformations a community can undergo: \textit{emerge} vs. \textit{dissolve} (a community appears vs. disappears), \textit{grow} vs. \textit{shrink} (a community size increases vs. decreases), \textit{split} vs. \textit{merge} (a community breaks down into smaller ones vs. several communities form a larger one). Note that these events appear under different names in the literature, e.g. \textcite{Palla2007} call them birth, death, growth, contraction, merging and splitting, respectively. 

The evolution of the community structure can be characterized by tracking and counting these events, but some authors also have used them to define more synthetic measures. Toyoda \& Kitsuregawa define their evolution metric as rates of changes, by focusing on the different types of changes: growth rate (total increase in nodes per unit time), stability (total change in nodes), novelty (increase due to node creation), disappearance rate (decrease due to node deletion), merge rate (increase in nodes due to merges) and split rate (decrease in nodes due to splits). On the same principle, \textcite{Asur2009} define their popularity index as the difference between the numbers of nodes joining and leaving a community of interest, for a given period. 

A number of measures have been proposed to assess the stability of the graph or of various of its parts, through the expression of some form of auto-correlation. \textcite{Palla2007} monitor the stability of a community through their \textit{stationarity measure}, based on the averaging of a custom auto-correlation function processed from the community appearance to its disappearance. This auto-correlation function is related to Jaccard's coefficient, i.e. it is based only on the evolution of the nodes community \textit{membership}, and ignores completely the link distribution. On the same principle Clauset \& Eagle have proposed a measure designed for the whole graph, and taking the topology into account: their \textit{adjacency correlation} measures how similar, topologically speaking, two consecutive time slices are \parencite{Clauset2007}. \textcite{Braha2009} apply a similar approach, except they compare all pairs of time slices (not just consecutive ones). On the same note, \textcite{Bost2016} uses Jaccard's coefficient to compare consecutive node neighborhoods in a dynamic network, and segments the neighborhood evolution into meaningful distinct parts.

Let us now stress how the method presented in this article is positioned relatively to this existing body of work. First, it is not an adaptation of some existing measure: the method was designed specifically to take advantage of dynamic networks. Second, it is a microscopic approach, because it focuses on a node of interest and its neighborhood, through the so-called \textit{ego-network} of this node. But it also borrows from the mesoscopic methods based on community events, because it adapts them to the context of the node neighborhood. Our idea is to characterize the evolution of a node through changes in its direct neighborhood, described by the occurrence of so-called \textit{neighborhood events}. Finally, our method also relates to the work of Desmier \textit{et al}. as well as to our previous work \parencite{Orman2014a} in that we use sequential pattern mining to extract meaningful information from the identified sequences of neighborhood events. Note that such pattern mining is just one possible post-analysis method among others, though.

\section{Methods}
\label{sec:Methods}
In  this section, we start with introducing the concept of node neighborhood evolution events (\ref{sec:neighborhoodevents}), before explaining how these can be detected and used (\ref{sec:EventAnalysis}).

\subsection{Neighborhood Evolution Events}
\label{sec:neighborhoodevents}
\begin{figure}[t]
	\center
	\includegraphics[width=1\textwidth]{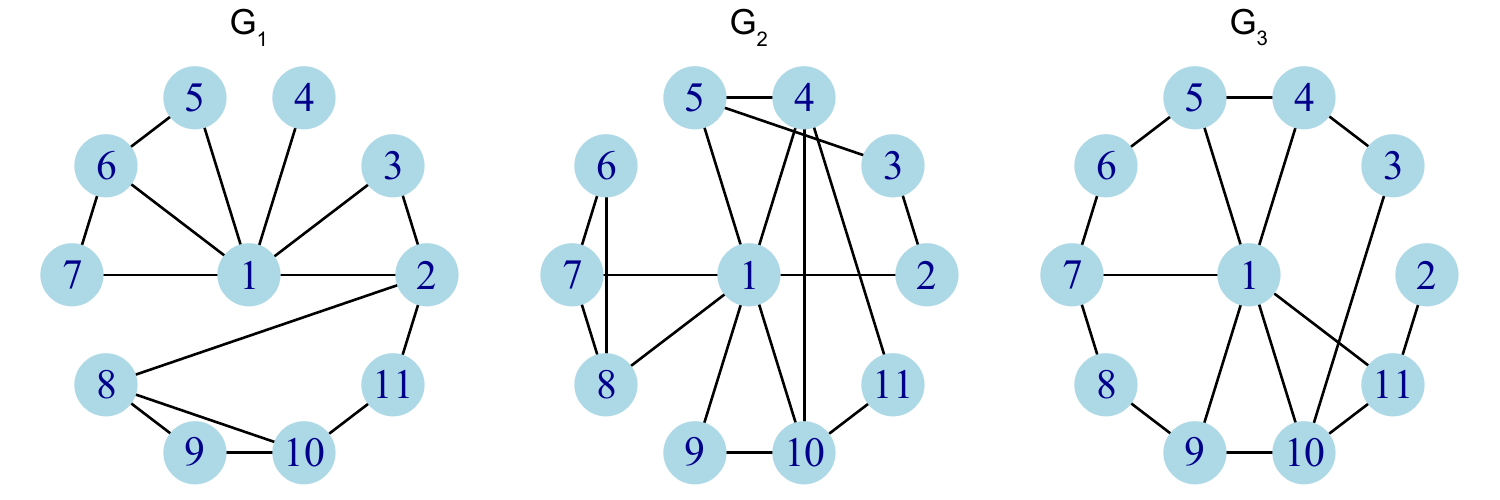}
	\caption{Example of dynamic network containing three time slices.}
	\label{fig:toyExample}
\end{figure}

This subsection is twofold: we first explain certain preliminary concepts and notations, and then properly define neighborhood evolution events. We illustrate our explanations with the toy example network from Figure~\ref{fig:toyExample}.

\subsubsection{Preliminary definitions}
\label{sec:prelim}
We start by properly defining the type of dynamic network used in this article, before deriving \textit{ad hoc} concepts used in the rest of this section to define neighborhood events.

\begin{definition}[Dynamic network and Time slice]
	\label{def:DynamicNet}
    A \textit{Dynamic network} $G$ is a sequence of $\theta$ chronologically ordered time slices $G_t$ ($1\leq t\leq\theta$):
    \begin{equation}
    	G = \langle G_1, \ldots, G_\theta \rangle
    \end{equation}
    
    A \textit{time slice} is a static graph defined as $G_t = (V,E_t)$, where $V$ is the set of nodes and $E_t \subseteq V \times V$ is the set of links.
\end{definition}

The set of links can vary from one time slice to the other, but the set of nodes is the same for all time slices, and we can consequently note $n = |V|$ the size of each time slice (and of the dynamic network as well). In Figure~\ref{fig:toyExample}, we describe a dynamic network $G=\langle G_1, G_2, G_3 \rangle$, in which we have $\theta=3$ and $n=11$. Remark that the number of nodes does not change with time, whereas the number of links and their distribution do. 

\begin{definition}[Ego-network]
	\label{def:EgoNet}
	The \textit{ego-network} (or egocentric network) of node $v \in V$ at time $t$ ($1\leq t\leq\theta$) is the subgraph noted $G_t(v) = (V_t(v),E_t(v))$ and such that:
    \begin{align}
    	V_t(v) &= \{v\} \cup N_t(v) \\
        E_t(v) &= \{(u_i,u_j) \in E_t: u_i,u_j \in V_t(v)\}
    \end{align}
where $N_t(v)$ is the first order neighborhood of $v$ at time $t$.
\end{definition}

\begin{figure}[t]
	\center
	\includegraphics[width=1\textwidth]{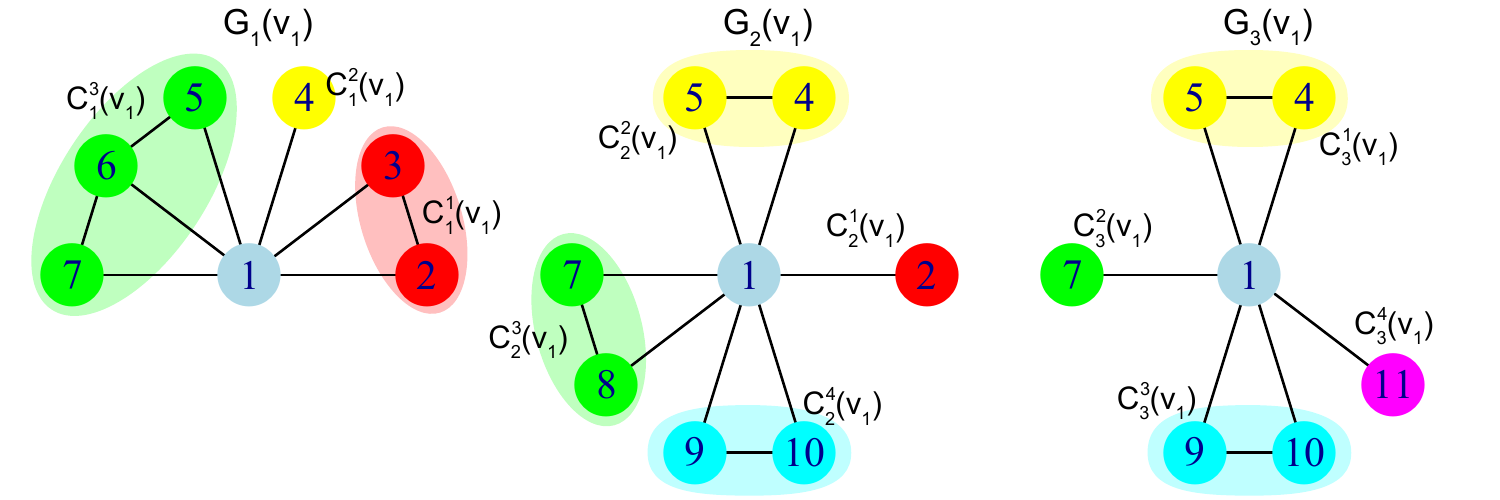}
	\caption{Ego-networks extracted from Figure~\ref{fig:toyExample}, for node $v_1$.}
	\label{fig:toyExampleEgo}
\end{figure}

Put differently, the ego-network $G_t(v)$ is the induced subgraph of $G_t$ constituted of node $v$ and its neighborhood in $G_t$. In Figure~\ref{fig:toyExampleEgo}, we represent the evolution of the ego-network of node $v_1$ from Figure~\ref{fig:toyExample}, for three time slices $t=\{1,2,3\}$. This results in three consecutive subnetworks $G_1(v_1)$, $G_2(v_1)$ and $G_3(v_1)$. Note that, at each time slice, both the numbers of nodes and links can change depending on the neighborhood of node $v_1$. 

\begin{definition}[Diminished ego-network]
	\label{def:DiminishedEgoNet}
    The \textit{diminished ego-network} $G'_t(v) = (V'_t(v), E'_t(v))$ is the subgraph obtained by removing $v$ and all its attached links from the ego-network $G_t(v)$. It is such that:
	\begin{align}
		V'_t(v) &= N_t(v) \\
		E'_t(v) &= \{(u_i,u_j) \in E_t: u_i,u_j \in V'_t(v)\}
	\end{align}
\end{definition}

The diminished ego-network is the induced subgraph of $G_t$ for $N_t(v)$. In Figure~\ref{fig:toyExampleEgo}, it corresponds to the subgraphs obtained when removing $v_1$.

\begin{definition}[Ego-component and Neighborhood partition]
	\label{def:EgoCompNeighPartition}
	Let $G'_t(v) = (V'_t(v), E'_t(v))$ be the diminished ego-network of $v \in V$ at time $t$. This graph is composed of $\pi_t(v)$ components (i.e. maximal connected subgraphs) called \textit{ego-components} and noted $C_t^i(v) \subseteq N_t(v)$.
    
    The set $\mathcal{C}_t(v) = \{ C_t^1(v),..., C_t^{\pi_t(v)}(v) \}$ of these $\pi_t(v)$ ego-components constitutes a partition of the original neighborhood of the node, called the \textit{neighborhood partition}.
\end{definition}

In practice, a diminished ego-network is often constituted of several components, i.e. $\pi_t(v)>1$. These ego-components constitute a partition of $N_t(v)$, so by definition we have $\bigcup_i C_t^i(v) = N_t(v)$ and $\bigcap_i C_t^i(v) = \emptyset$. In Figure~\ref{fig:toyExampleEgo}, the nodes belonging to the same ego-component are represented with the same color. At time $t=1$, for instance, there are $\pi_1(v_1)=3$ ego-components: $C_1^1(v_1) = \{v_2,v_3\}$ (red), $C_1^2(v_1) = \{v_4\}$ (yellow) and $C_1^3(v_1) = \{v_5,v_6,v_7\}$ (green). Therefore, the neighborhood partition of $v_1$ at time $t=1$ is $\mathcal{C}_1(v_1) = \{C_1^1(v_1),C_1^2(v_1),C_1^3(v_1)\}$. 


\begin{definition}[Event matrix]
	\label{def:EventMatrix}
	Let $\mathcal{C}_t(v)$ and $\mathcal{C}_{t+1}(v)$ ($t<\theta$) be two consecutive neighborhood partitions for some node $v$. The \textit{event matrix} $M_{t,t+1}(v)$ is the $\pi_t(v) \times \pi_{t+1}(v)$ matrix whose elements are defined as follows: 
    \begin{equation}
    	m_{t,t+1}^{ij}(v) = |C_t^i(v) \cap C_{t+1}^j(v)|
    \end{equation}
with $1 \leq i \leq \pi_t(v)$ and $1 \leq j \leq \pi_{t+1}(v)$).
\end{definition}

In other words, this neighborhood matrix is the confusion matrix allowing to compare both considered partitions: each element $m_{t,t+1}^{ij}(v)$ contains the number of nodes belonging to the $i^{th}$ ego-component at time $t$ \textit{and} to the $j^{th}$ one at $t+1$. In our toy example from Figure~\ref{fig:toyExampleEgo}, we get the following event matrices for node $v_1$:
\begin{equation}
	M_{1,2}(v_1) =
    \begin{bmatrix}
		1 & 0 & 0 & 0 \\
		0 & 1 & 0 & 0 \\
		0 & 1 & 1 & 0 
	\end{bmatrix}
	~;~
	M_{2,3}(v_1) =
	\begin{bmatrix}
		0 & 0 & 0 & 0 \\
		2 & 0 & 0 & 0 \\
		0 & 1 & 0 & 0 \\
		0 & 0 & 2 & 0 \\
	\end{bmatrix}
    \label{eq:ToyMatrices}
\end{equation}

\subsubsection{Notion of Neighborhood Event}
\label{sec:evolution}
Before defining the events themselves, we need first to introduce the notion of identical partitions.

\begin{definition}[Identical partitions]
	Two consecutive neighborhood partitions $\mathcal{C}_t(v)$ and $\mathcal{C}_{t+1}(v)$ are \textit{identical} if the following conditions are all met:
    \begin{align}
    	\forall i, \exists j, \forall j' \neq j: m_{t,t+1}^{ij}(v) \neq 0 \land m_{t,t+1}^{ij'}(v) = 0, \\
    	\forall j, \exists i, \forall i' \neq i: m_{t,t+1}^{ij}(v) \neq 0 \land m_{t,t+1}^{i'j}(v) = 0, \\
   		\sum_{i,j} m_{t,t+1}^{ij}(v) = |N_t(v) \cup N_{t+1}(v)|.
    \end{align}
\end{definition}

These partitions are thus considered identical if each column and each row of $M_{t,t+1}(v)$ contains exactly one non-zero value, and if the sum over $M_{t,t+1}(v)$ is equal to the number of nodes which are neighbors of $v$ at $t$ \textit{or} $t+1$. In other words, they are not identical if some neighbor appears, disappears, or switches ego-component during the considered interval.

If two consecutive neighborhood partitions are not identical, we consider that at least one \textit{evolution event} is happening. We define six types of such events: \textit{birth} vs. \textit{death}, \textit{merge} vs. \textit{split}, and \textit{expansion} vs. \textit{contraction}. Note that it is possible for an ego-component to simultaneously undergo several events. The matrices described in Eq. (\ref{eq:ToyMatrices}) fail to fulfill our conditions, so the partitions considered in our toy example are not identical, and evolution events are consequently taking place.

\begin{definition}[Birth and Death events]
	The ego-component $C_{t+1}^j(v)$ is \textit{born} at time ${t+1}$ iff $\sum_i m_{t,t+1}^{ij}(v) = 0$. This amounts to having a column of zeros in the event matrix.
    
    Symmetrically, $C_t^i(v)$ \textit{dies} at time $t+1$ iff $\sum_j m_{t,t+1}^{ij}(v) = 0$. This amounts to having a row of zeros in the event matrix.
\end{definition}

We say there is a birth event if a new ego-component appears from one time slice to the next one, i.e. it contains only nodes not belonging to any ego-component of the previous time slice. In our example, if we consider $M_{1,2}(v_1)$, we see a birth taking place for $C_2^4(v_1)$, because column $j=4$ contains only zeros. In Figure~\ref{fig:toyExampleEgo}, this corresponds to the apparition of the cyan ego-component (nodes $v_9$ and $v_{10}$) at $t=2$.

Symmetrically, a death event appears if an ego-component disappears from one time slice to the next one, i.e. all the nodes it contains in the previous time slice do not appear in any ego-component of the current one. In our example, if we consider $M_{2,3}(v_1)$, we observe a death occurring for $C_2^1(v_1)$, since row $i=1$ contains only zeros. In Figure~\ref{fig:toyExampleEgo}, this corresponds to the disappearance of the red ego-component (node $v_2$) from $t=2$.

\begin{definition}[Merge and Split events]
	The ego-component $C_{t+1}^j(v)$ is the result of a \textit{merge} at time $t+1$ iff $|\{ i: m_{t,t+1}^{ij}(v) > 0\}| > 1$. This amounts to having more than one non-zero value in the $j^{th}$ column of the event matrix.

	Symmetrically, $C_t^i(v)$ is \textit{split} at time $t+1$ iff $|\{ j: m_{t,t+1}^{ij}(v) > 0\}| > 1$. This amounts to having more than one non-zero value in the $i^{th}$ row of the event matrix.
\end{definition}

A merge event corresponds to the combination of two or more ego-components into a single one, i.e. the resulting ego-component contains nodes coming from at least two distinct ego-components from the previous time-slice. In our example, if we consider $M_{1,2}(v_1)$, we observe a merge resulting in $C_2^2(v_1)$, since column $j=2$ contains two ones. In Figure~\ref{fig:toyExampleEgo}, this corresponds to the fusion of the yellow ego-component (node $v_4$) and a part of the green ego-component (node $v_5$) from $t=1$, to form the yellow ego-component at $t=2$.

Symmetrically, a split event occurs if an ego-component breaks down into two or more ego-components, i.e. some of its nodes belong to different ego-components in the next time slice. In our example, if we consider $M_{1,2}(v_1)$, $C_3^1(v_1)$ undergoes a split, since row $i=3$ contains two ones. In Figure~\ref{fig:toyExampleEgo}, this corresponds to the split of the green ego-component, whose node $v_5$ joins node $v_4$ to form the yellow ego-component at $t=2$, while $v_7$ becomes a part of a new green ego-component together with $v_8$.

\begin{definition}[Expansion and Contraction events]
	The ego-component $C_{t+1}^j(v)$ \textit{expands} at time $t+1$ iff $\sum_i m_{t,t+1}^{ij}(v) < |C_{t+1}^j(v)|$. This amounts to having more nodes in $C_{t+1}^j(v)$ than in the $j^{th}$ column.

	Symmetrically, $C_t^i(v)$ \textit{contracts} at time $t+1$ iff $\sum_j m_{t,t+1}^{ij}(v) < |C_t^i(v)|$. This amounts to having more nodes in $C_{t+1}^i(v)$ than in the $i^{th}$ row.
\end{definition}

There is an \textit{expansion} event if an ego-component includes new nodes from one time slice to the next, i.e. if it contains at least one node which did not belong to any ego-component in the previous time slice. This is the case when the size of the ego-component (at $t+1$) is larger than the sum over its corresponding column in the event matrix. In our example, if we consider $M_{1,2}(v_1)$, an expansion occurs for $C_2^3(v_1)$, because the sum over column $j=3$ is $1$, which is smaller than the ego-component size $|C_2^3(v_1)|=2$. In Figure~\ref{fig:toyExampleEgo}, this corresponds to the fact the green ego-component gets a new node $v_8$ at $t=2$. Note that the same ego-component also loses two nodes ($v_5$ and $v_6$) due to other events (a merge and a contraction, respectively) occurring simultaneously.

Symmetrically, a \textit{contraction} event happens if some nodes of an ego-component disappear from the neighborhood from one time slice to the next, i.e. if at least one of its nodes from the previous time slice does not belong to any ego-component in the current one. This is the case when the size of the ego-component (at $t$) is larger than the sum over its corresponding row in the event matrix. In our example, if we consider $M_{1,2}(v_1)$, a contraction occurs for $C_1^3(v_1)$, because the sum over row $i=3$ is $2$, which is smaller than the ego-component size $|C_1^3(v_1)|=3$. In Figure~\ref{fig:toyExampleEgo}, this corresponds to the fact node $v_6$ of the green ego-component disappears from the neighborhood at $t=2$.

The events we use are quite similar to those proposed in several works studying the evolution of dynamic community structures, such as those of \textcite{Toyoda2003} or \textcite{Palla2007}. However, their semantics is quite different, since in our case these events are experienced by \textit{node neighborhoods}, whereas in the other works they concern \textit{communities}. The main consequence is that in our case, it is possible for a node to simply disappear from the neighborhood, or appear into it. On the contrary, in the context of community structures, the whole network is considered, so nodes can only switch communities (provided the communities form a partition of $V$, and $V$ is fixed). All the events considered by Toyoda \& Kitsuregawa or Palla \textit{et al}. are different types of such switches. By comparison, in our case the birth and expansion (resp. death and contraction) events are based only on nodes entering (resp. leaving) the neighborhood. Only the merge and split events are based on nodes already present and remaining in the neighborhood.

Finally, as mentioned before, and as illustrated by our example, several events can occur simultaneously. Their occurrence and frequency depends on the dynamics of the network structure in the neighborhood of the considered node. This is why we make the assumption that the sequence of these events can be used to characterize this node.

\subsection{Processing of Neighborhood Events}
\label{sec:EventAnalysis}
In this subsection, we first describe our algorithm to identify and extract neighborhood events from a dynamic network and constitute a sequential database, then explain how sequential pattern mining can be used to extract relevant information from this database.

\subsubsection{Event Detection}
\label{sec:EventDetection}
We now need to identify all the evolution events taking place between each pair of consecutive time slices, for each node. This procedure is described as Algorithm~\ref{code:egoComm}, and outputs a database noted $S$. It takes the form of a set of $n$ sequences, each one containing the ordered events underwent by the neighborhood of a given node (sequence-related concepts are described in more details in Section \ref{sec:EventPatterns}).

Each node is processed iteratively (first loop), as well as each pair of consecutive time slices (second loop). We first extract the ego-networks of the considered node $v$ for both considered time slices $t$ and $t+1$, using the function $extractEgoNetwork$ (lines \ref{algo:egoNetT0} and \ref{algo:egoNetT1}). This function performs a breadth-first search (BFS) restricted to $v$ and its neighbors. Let us note note $k_t(v)$ the degree of $v$ at time $t$. In the worst case, $v$ and its neighborhood form a clique, which means the ego-network contains $k_t(v)(k_t(v)-1)/2$ links. Since the complexity of the BFS is linear in the number of nodes and links, we get $O((k_t(v)+1) + k_t(v)(k_t(v)-1)/2) = O(k(v)^2)$ for our function.

\begin{algorithm}
	\caption{Finding Evolution Events}
	\label{code:egoComm}
	\begin{algorithmic}[1]
		\REQUIRE { $G$}
		\ENSURE{ $S$ }
		\FOR {$v \in V $ }
			\FOR {$t \in \{1 \ldots (\theta-1)\}$} 
				\STATE {$G_t(v) \leftarrow extractEgoNetwork(G_t, v)$} \label{algo:egoNetT0}
				\STATE {$G_{t+1}(v) \leftarrow extractEgoNetwork(G_{t+1}, v)$} \label{algo:egoNetT1}
				\STATE {$G'_t(v) \leftarrow removeNode(G_t(v), v)$} \label{algo:removeT0}
				\STATE {$G'_{t+1}(v) \leftarrow removeNode(G_{t+1}(v),v)$} \label{algo:removeT1}
				\STATE {$\mathcal{C}_t(v) \leftarrow identifyEgoComponents(G'_t(v))$} \label{algo:egoCompT0}
				\STATE {$\mathcal{C}_{t+1}(v) \leftarrow identifyEgoComponents(G'_{t+1}(v))$} \label{algo:egoCompT1}
				\STATE {$M_{t,t+1}(v) \leftarrow processEventMatrix(\mathcal{C}_t(v),\mathcal{C}_{t+1}(v))$} \label{algo:eventMat}
				\STATE {$s_v \leftarrow detectEvents(M_{t,t+1}(v))$} \label{algo:detEvents}
				\STATE {$S \leftarrow insertSequence(S, s_v)$} \label{algo:insertSeq}
			\ENDFOR
		\ENDFOR
	\end{algorithmic}
\end{algorithm}

The next step consists in applying function $removeNode$ (lines \ref{algo:removeT0} and  \ref{algo:removeT1}), whose role is to delete $v$ from both $G_t(v)$ and $G_{t+1}(v)$, in order to get the corresponding diminished ego-networks. For time slice $t$, this is simply done by removing $1$ node and $k_t(v)$ links, i.e. $O(k_t(v))$ operations. 
Both diminished ego-networks are then fetched to function $identifyEgoComponents$ (lines \ref{algo:egoCompT0} and \ref{algo:egoCompT1}), in order to process the corresponding neighborhood partitions. For time slice $t$, this is achieved by iteratively performing a BFS until all nodes in $G'_t(v)$ have been explored, so we have a complexity in $O(k_t(v)^2)$ again.

Function $processEventMatrix$ is then applied to get the event matrix (line \ref{algo:eventMat}). We basically have to process the intersection cardinality for all pairs of ego-components in partitions $\mathcal{C}_t(v)$ and $\mathcal{C}_{t+1}(v)$. Computing the intersection of $C_t^i(v)$ and $C_{t+1}^j(v)$ requires $O(|C_{tv}^{i}| + |C_{(t+1)v}^{j}|)$ operations if both sets are ordered. The worst case is when both partitions contain only singletons, leading to a complexity in $O(k_t(v)k_{t+1}(v))$. This can be simplified to $O(k_t(v)^2)$ if we suppose the degree of a node is relatively stable. 
We then look for events in this matrix, through function $detectEvents$ (line \ref{algo:detEvents}). It checks the conditions given for each type of event in section ~\ref{sec:evolution}. All of them require to scan the whole matrix, leading again to a complexity in $O(k_t(v)k_{(t+1)}(v)) \approx O(k_t(v)^2)$ in the worst case (partitions containing only singletons). 
Finally, all detected events are added to the database thanks to function $insertSequence$ (line \ref{algo:insertSeq}), which operates in constant time.

The total computation for a node $v$ and a pair of consecutive time slices is in $O(k_t(v)^2)$. Instead of $k_t(v)$, let us consider the degree averaged over all nodes in all time slices $\langle k \rangle = 
2m/(\theta n)$, where $m = \sum_t |E_t|$ ($n$ being the number of nodes in each time slice). The complexity can then be rewritten $O(m^2/(\theta n)^2)$. By taking both loops into account, the expression becomes $O(m^2/(\theta n)^2 \cdot \theta n) = O(m^2/(\theta n))$. 
The algorithm has a computational complexity in $O(m^2/(\theta n))$, where $m = \sum_t |E_t|$, $n$ is the number of nodes in each time slice, and $\theta$ is the number of time slices.
Note that each node in each time slice can be treated independently, so this computation can be parallelized, which significantly speeds it up in practice. 

\subsubsection{Sequential Pattern Mining}
\label{sec:EventPatterns}
As mentioned before, we see evolution events as categorical variables describing the nodes. Once our database is constituted, we can analyze them through sequential pattern mining tools. Sequential pattern mining is the data mining task aiming at discovering frequent subsequences in sequentially structured databases \parencite{Mabroukeh2010}. This approach was previously used in network science for interpreting communities in dynamic attributed networks \parencite{Orman2015}. In this work, we use it as a descriptive tool to find out general trends of node evolution. Here, we shortly explain the related concepts in the perspective of our work and how we use them. 

In our work, an \textit{item} $l_i$ is an evolution event which can be \textit{birth}, \textit{death}, \textit{merge}, \textit{split}, \textit{expansion} or \textit{contraction}. 
The set of all items is noted $I$. An \textit{itemset} $h$ is any subset of $I$. Although itemsets are sets, in the rest of this article we represent them between parentheses, e.g. $h=(l_1,l_3,l_4 )$ because it is the standard notation in the literature. 

A \textit{sequence} $s= \langle h_1,\cdots , h_m \rangle $ is a chronologically ordered list of itemsets. Two itemsets can be consecutive in the sequence while not correspond to consecutive time slices: the important point is that the first to appear must be associated to a time slice preceding that of the second one. In other words, $h_i$ occurs before $h_{i+1}$ and after $h_{i-1} $. The \textit{size} of a sequence is the number of itemsets it contains. A sequence $\alpha=\langle a_{1},\ldots,a_{\mu} \rangle$ is a \textit{sub-sequence} of another sequence $\beta=\langle b_{1},\ldots,b_{\nu}\rangle$ iff $\exists i_{1},i_{2},\ldots,i_{\mu}$ such that $1\leq i_{1}<i_{2}<\ldots<i_{\mu}\leq \nu$ and $a_{1}\subseteq b_{i_1},a_{2}\subseteq b_{i_2}, \ldots,a_{\mu}\subseteq b_{i_{\mu}}$. This is noted $\alpha\sqsubseteq\beta$. It is also said that $\beta$ is a \textit{super-sequence} of $\alpha$,which is noted $\beta\sqsupseteq\alpha$. 

The \textit{node sequence} $s_v$ of a node $v$ is a specific type of sequence of size $\theta-1$. We have $s_v=\langle(l_{11},\dots,l_{k1} ) \dots (l_{1(\theta-1)},\dots,l_{k(\theta-1)} ) \rangle$, where $l_{it}$ is the item representing the occurrence of a $i^{th}$ evolution event in the neighborhood of $v$ between time slices $t$ and $t+1$. The \textit{sequence database} $S$ built in the previous subsection is the collection of node sequences $s_v$ for all nodes in the considered network. The set of \textit{supporting nodes} $\mathcal{S}(s)$ of a sequence $s$ is defined as $\mathcal{S}(s) =\{v\in V:s_v \sqsupseteq s\} $. The support of a sequence $s$, $Sup(s) = |\mathcal{S}(s)|/n$ , is the proportion of nodes, in $G$, whose node sequences are equal to $s$, or are super-sequences of $s$.

Sequences can be mined based on a number of different constraints, as illustrated in the sequential pattern mining literature \parencite{Mabroukeh2010}. We take advantage of the SPMF framework \parencite{Fournier-Viger2010}, which contains several sequential pattern mining algorithms, to detect different types of patterns. Finding the most interesting patterns is an important and difficult issue. Here, without deepening this subject, we focus on frequent patterns with additional constraints on closeness and length. First, thanks to the Clospan algorithm \parencite{Yan2003}, we search for \textit{closed frequent sequential patterns} (CFS). In this context, a sequence $s$ is considered as frequent (FS) if its support is larger than a predefined minimum support threshold $min_{sup}$, i.e. $Sup(s) \geq min_{sup}$. A CFS is a FS with no super-sequence possessing the same support. We start with the maximal value ($1.0$) for minimum support and iteratively decrease it with a $0.1$ step, until completion. Second, we take advantage of the TKS algorithm \parencite{Fournier-Viger2013}, in which the notion of FS is expressed in terms of $k$ most supported patterns. This algorithm allows putting constraints on the minimal/maximal length of the detected sequences. We look for the \textit{longest frequent sequences} (LFS) by setting the minimal length parameter to the longest value treatable by our hardware. The length of a sequence gives us an idea about the duration of a trend, but longer patterns have also smaller support. In our case, nodes following LFSs may have interesting topological properties or a specific role in the network.

Depending on the homogeneity of the considered dataset, it may not be possible to find interesting patterns when considering the whole network. In this case, one can mine subparts of the network separately, as was done in \parencite{Orman2015} for communities. We apply this principle using clusters identified based on the sequences of neighborhood events. For this purpose, we describe each node by a sequence of numerical values, each one representing the number of neighborhood events it undergoes at one time slice. We process the distance between two nodes through \textit{Dynamic Time Warping} (DTW) \parencite{Toni2009}, which was designed to measure the distance between two numerical time series. Once we have the distance between each pair of nodes, we apply a standard hierarchical clustering method, which outputs a dendrogram. We select its best cut using the average \textit{Silhouette} width (ASW) \parencite{Rousseeuw1987}. The Silouhette width measures how much a node fits its cluster. It ranges from $-1$ (worst fit) to $+1$ (best fit). When averaged over a dataset, it allows assessing the quality of the considered partition. We then look for both types of patterns in each cluster separately. In order to select the most relevant ones, we take advantage of an additional criterion called the \textit{growth rate}. The growth rate of a sequential pattern $s$ relatively to a cluster $C$ is $Gr(s,C) = Sup(s,C) / Sup(s,\overline{C})$. Here, $\overline{C}$ is the complement of  $C$ in $V$, i.e. $ C=V\setminus\overline{C} $. The growth rate measures the emergence of $s$. It is the ratio of the support of $s$ in  $C$ to the support of $s$ in $\overline{C}$. Thus, a value larger than $1$ means $s$ is particularly frequent (i.e. emerging) in  $C$, when compared to the rest of the network.

The cluster analysis we perform can be considered as complementary to the pattern mining method. Indeed, both rely on the analysis of event-based sequences, but the former focuses on the numbers of events occurring, without regards for their types, whereas the latter depends only on the presence/absence of types of events, and not on their numbers.

\section{Experimental validation}
\label{sec:results}
In this section, we first describe the data on which we applied our tools (Subsection \ref{sec:DataDescr}). Second, we give the results obtained in terms of neighborhood events, discuss their distribution over time, and propose an interpretation for the studied systems (Subsection \ref{sec:EventDistrib}). Third, we focus on the patterns extracted from these sequences of events, and discuss them relatively to the systems (Subsection \ref{sec:sequential}). Fourth, we turn to cluster analysis and describe and comment the obtained clusters in terms of patterns (Subsection \ref{sec:ClusterAnalysis}).

\begin{table}[b]
	\caption{Real-world networks used in the experimental evaluation}
	\label{table:networks}
	\centering
	\begin{tabular}{ |p{1.6cm}  | r| r| r | r| }
		\hline
		Network & Nodes & Active nodes & Time slices & Time span \\ \hline
		DBLP & 2145 & 2046 & 10 & 1990-2012 \\ \hline
		LastFM & 1701 & 1269 & 10 & Jan-Dec 2013 \\ \hline
		Enron & 28802&  28649 & 46  & 1997-2002\\
		\hline
	\end{tabular}
\end{table}

\begin{figure*}[tb]
	\center
	\includegraphics[width=1\textwidth]{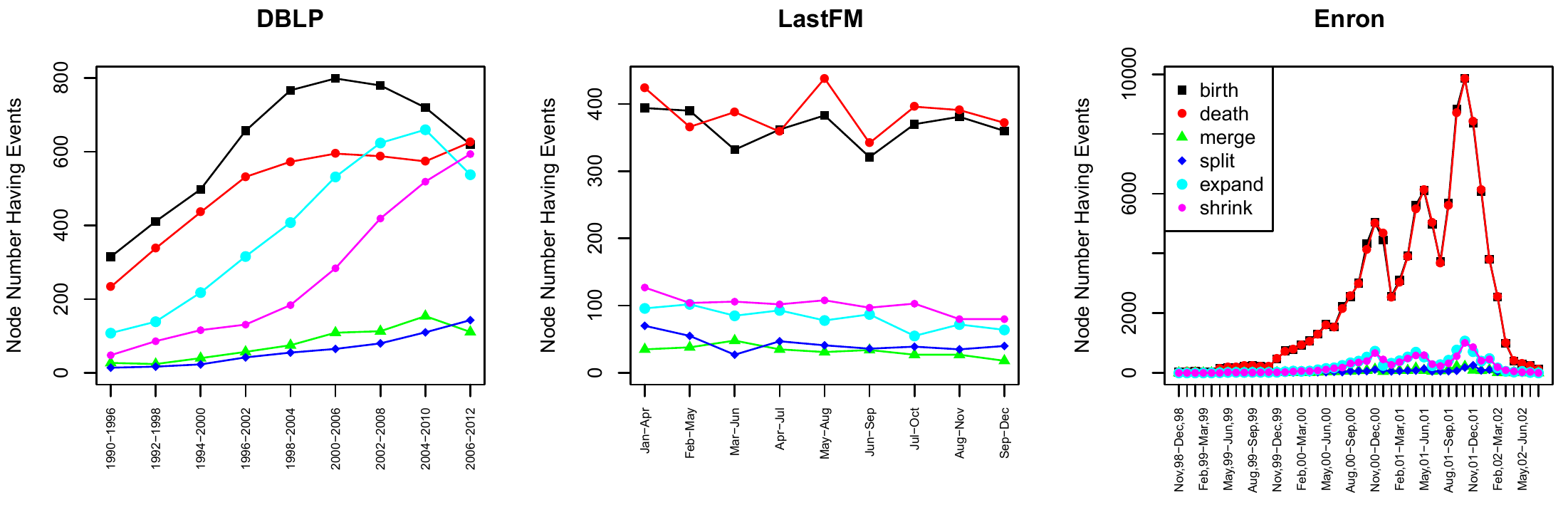}
	\caption{Number of nodes undergoing each type of event, per time slice, for all three considered dynamic networks}
	\label{fig:eventNodeNumber}
\end{figure*}

\subsection{Data Description}
\label{sec:DataDescr}
We have studied $3$ dynamic networks based on three different datasets: DBLP, LastFM and Enron. Both the DBLP and LastFM networks were previously used in the literature: the former to detect evolution patterns \parencite{Desmier2012}, and the latter to interpret communities through sequential pattern mining \parencite{Orman2015}. The Enron network was specifically extracted for this study, as described below. A summary of the networks can be found in Table~\ref{table:networks}: number of nodes, number of nodes undergoing at least one event, number of time slices, and time span of the considered data.

When extracting slowly evolving dynamic networks from raw data (by opposition to streaming networks, as described in the introduction), one of the most important point is the time scale used to discretize or bin the original data. In other words, it is the duration represented by each time slice. This point has been discussed in the literature, and it has been shown this parameter can have significant effect on the network topological properties \parencite{Clauset2007, Krings2012, Sulo2010}. The DBLP and LastFM networks have been the object of previous studies, so we assumed the problem of selecting the most appropriate time scale had already been solved for them, and used them as is. For the Enron network, we had access to the clean raw data and took advantage of this to apply the method proposed in \parencite{Clauset2007} in order to identify the most appropriate time scale.

\subsubsection{DBLP Network}
The DBLP dynamic network considered in this work was first used in \parencite{Desmier2012}. It is a co-authorship network extracted from the DBLP database. Each one of its $2145$ nodes represents an author working in one or several Computer Science subfields among: Databases, Data Mining, Knowledge Discovery, Information Retrieval and Artificial Intelligence. Two nodes are connected if the corresponding authors have published an article together at the considered time, in a selected set of $43$ conferences and journals. Only authors who have at least $10$ publications between 1990 and 2012 are included in the network. Each time slice corresponds to a period of five years. There are a total of $10$ time slices ranging from 1990 to 2012. The consecutive periods have a three year overlap for the sake of stability.

\subsubsection{LastFM Network}
The LastFM dynamic network considered in this work was first used in \parencite{Orman2015}. It represents the music tastes and friend relationships of $1701$ LastFM users registered to the \textit{Jazz} group on this musical platform. Each node corresponds to a user, and two users are connected by a link if two conditions are fulfilled: 1) the users are friends on the LastFM platform, and 2) both of them have listened to the music of at least one common artist over the considered period of time. The network represents the evolution of the users in the year 2013. The time slices were one month long in the first published occurrence of this network \parencite{Orman2015}, however in this article we used a three month period with two months overlap to ensure continuity between consecutive times slices. Hence, we have $10$ time slices in total. This new version is publicly available online\footnote{\url{https://dx.doi.org/10.6084/m9.figshare.3203404}}.

\subsubsection{Enron Network}
Enron is a well-known dataset, which has been widely studied in network science and text mining. In particular, it appears in the literature under the form of several distinct static and dynamic networks. In this work, we perform another extraction based on the original dataset\footnote{\url{https://www.cs.cmu.edu/~./enron/}}, in order to get a network appropriate to the application and evaluation of our analysis method. As mentioned before, we used the approach of Clauset \& Eagle to identify the best time scale \parencite{Clauset2007}. Put briefly, in order not to leave the scope of this article: we compared the networks obtained with one- (without overlap), two- (without overlap), three- (without and with a 1-month overlap), six- (without and with 1- and 2-month overlaps) and twelve-month time slices (without, with 1-, 2- and 3-month overlaps) and identified 1 month without any overlap as the most appropriate time scale. Our network is publicly available online\footnote{\url{https://dx.doi.org/10.6084/m9.figshare.3203410}}.

It contains $158$ nodes representing Enron employees between $1997$ and $2002$. All the addresses in the \texttt{From} and \texttt{To} fields of each email are considered, resulting in a network of $28802$ nodes representing distinct email addresses. Two nodes are connected if the corresponding persons emailed each other during the given time slice. Note that the networks extracted from such datasets are usually directed. As this information was not relevant to our studies at this time, we did not make any distinction between sender and receiver, and thus produced an undirected dynamic network. Since we used a one-month long time slice, this dynamic network is composed of $46$ time slices.

\subsection{Event Distribution}
\label{sec:EventDistrib}
We identified the neighborhood events by applying the method described in Section \ref{sec:EventDetection} to all three networks. Only a minority of nodes do not undergo a single event for the whole considered time span: $99$ for DBLP ($\simeq 4.6 \%$ of the network size), $432$ for LastFM ($\simeq 25 \%$) and $153$ for Enron  ($\simeq 0.5 \%$). Figure~\ref{fig:eventNodeNumber} displays the numbers of nodes undergoing the different types of events at each time slice. It is interesting to notice that the evolution of the event counts is very different from one network to the other. 
However, there are also some common points: births are the most common events for all the networks, followed by deaths, whereas the rarest ones are merges and splits.

This changes could be directly caused by evolution of the network structure, e.g. the densification process often observed in real-world systems could explain the increase in births. To check this assumption, we considered two topological measures: (1) the number of alive nodes (i.e. nodes having at least one neighbor) and (2) the link density. We studied the association between these measures and the number of events of each type, using Pearson's correlation coefficient and linear regression. It turns out there is no statistically significant linear relation between the studied measures and the numbers of events, for any of the three considered networks. According to these results, the occurrence of events is not directly due to the global changes in the network structure. We therefore suppose it is caused by other factors, which we discuss in the rest of this section.

\subsubsection{DBLP}
We observe an increase in all the events, starting from the beginning of the considered period ($1990$--$2012$). Then, around time slice $1994$--$2000$, there is a sudden increase for both births and expansions. The number of deaths also increases, but to a lesser extent. The increase in contractions occurs later, around time slice $2000$--$2006$. In comparison, splits and merges are relatively stable throughout the whole considered period. Moreover, until time slice $2004$--$2010$, there are more births, expansions and merges than deaths, contractions and splits, respectively. These observations indicate the monitored scientific communities underwent significant changes during this period. The high number of births can be interpreted as the apparition of new collaborations among researchers. The slower increase in deaths means these collaborations did not disappear right away and were likely to be relatively stable. The large increase in expansions shows authors tended to include new people into existing collaborations. The fact the number of contractions also largely increases, but only after a delay, indicates the initial expansion was followed by some adjustments to reduce the size of the collaborator groups.

Interestingly, the periods for which the plot exhibits a sudden increase of births and expansions ($1994$--$2000$ and $1996$--$2002$) correspond to the merger of two important Data Mining-related conferences, ECML and PKDD (occurring in $2001$) and the apparition of an important new one, ICDM (also in $2001$). ECML was more Machine Learning-oriented, whereas the focus of PKDD was more on Knowledge Discovery. The combination of these two conferences caused the encounter of many researchers coming from two different subfields, forcing them to share, discuss and exchange ideas. Although not the result of a merge, ICDM is one of the most important conferences on Data Mining, it has gathered numerous researchers from the whole world since its beginning. We believe the collaborations resulting from these academic events explain our observations regarding the neighborhood events in the DBLP network. Or, conversely, the events extracted from the network depict the historical evolution of the Data Mining field.

Time slice $2004$--$2010$ corresponds to a trend reversal: the numbers of births, expansions and merges plummet and are dominated by their counterparts (deaths, contractions and splits). Although it is a bit more far-fetched, this could correspond to the emergence of new domains relatively close to Data Mining, but possessing their own conferences. We think, in particular, about the large quantity of data produced by the Web 2.0 and its social media platforms, leading to the Big Data and Network Analysis fields. Under this assumption, the observed increase in deaths and splits could correspond to researchers starting to collaborate out of the (strictly speaking) Data Mining field, becoming less active in this domain and disappearing from our dataset.

\subsubsection{LastFM}
Like for DBLP, the most frequent events are births/deaths, followed by expansions/contractions, and then merges/splits. However, for LastFM the measured values are much stabler, and the plot does not display any sudden increase or drop. When checking the data, we observed most of these births/deaths do not concern large ego-components, but rather single nodes arriving/leaving the neighborhood. 

Our LastFM dataset corresponds to a $1$ year evolution of Jazz listeners, so the considered time span might not be long enough to uncover stronger changes in this network. Nevertheless, we obtained interesting results when characterizing local event evolution through sequential pattern mining, as shown in Section~\ref{sec:sequential}.

\subsubsection{Enron}
Like before, the most common event types are, by decreasing order: births/deaths, expansions/contractions and merges/splits. Symmetrical event types evolve jointly as in the other networks, but this is even more marked here. Since this is an email network, a birth and a death correspond to a conversation starting/restarting and ending/pausing, respectively. Put differently, if two persons stop sending each other emails for some time, a death event will occur, and if they then resume their exchange, it will be a birth. This is a common situation, which explains the much larger numbers of births and deaths. Expansions and contractions correspond to people joining or leaving existing conversations, which is much rarer. Merges and splits are even rarer because they happen when two conversations become one, or when some people start branching out of the conversation, which is quite unusual in email exchanges. 

Although it is not obvious in the plot due to the scale, all the types of events follow similar trends. First, there is a seasonal trend: sudden increases on the months of June and December. This could be due to the fact those months are particularly busy in the corporate sector, as they mark the first and second halves of the year. For instance, many reports describing the company situation are collectively prepared at these moments. Hence, it is natural that employees communicate more intensely, possibly with persons with which they do not usually work. The second trend goes through the years: the different types of exchanges gradually increase until the first month of $2002$, after which they drop. This date corresponds to the official declaration of bankruptcy of the company.

\subsection{Sequential Patterns of the Network}
\label{sec:sequential}
As explained in Section \ref{sec:EventPatterns}, we use the SPMF framework \parencite{Fournier-Viger2010} to identify the closed frequent sequences (CFS) and the longest frequent sequences (LFS) in the whole network. We now describe the main results obtained for the three considered datasets. Our goal here is not to be exhaustive, but rather to illustrate what can be done by combining our event-based approach and pattern mining methods.

\subsubsection{Closed Patterns} 
In Table~\ref{table:mostFreqPat}, we list the CFSs with the highest support rates for all three networks. For a pattern $s$, the last two columns correspond to the number of supporting nodes ($|\mathcal{S}(s)|$) and support rate ($Sup(s)$). Unsurprisingly (given the observations of Section \ref{sec:EventDistrib}), for all networks, the most frequent CFSs are composed only of births and/or deaths. Moreover, they are very short: one or two itemsets. All other highly supported patterns (not represented in the table) are very short ; longer patterns are much less supported. In the DBLP network, $95 \%$ of the nodes undergo at least one birth over the considered period, and the support rate is $94\%$ for a death. This means that between $1990$ and $2012$, the overwhelming majority of authors started a completely new collaboration and ended a collaboration (not necessarily the same). 

\begin{table}[!tb]
	\caption{Most supported Closed Sequential Patterns}
	\label{table:mostFreqPat}
	\centering
	\begin{tabular}{| l | p{4.6cm} | r | r |}
    	\hline
		Network & Sequential pattern $s$ & $|\mathcal{S}(s)|$ & $Sup(s)$ \\ 
        \hline
    	\multirow{2}{*}{DBLP} & $\langle (B)\rangle$  & 2041 & 0.95  \\
	     & $\langle (D)\rangle$  & 2032 & 0.94 \\ 
        \hline
	    \multirow{2}{*}{LastFM} &$\langle (D)\rangle$  & 1530 & 0.90  \\
	     & $\langle (B)\rangle$  & 1510 & 0.88  \\
        \hline
	    \multirow{2}{*}{Enron} & $\langle (B, D)\rangle$  & 28649 & 0.99  \\
	     & $\langle (B, D)$ $(B, D)\rangle$  & 28634 & 0.99 
        \\ \hline
	\end{tabular}
\end{table}

For LastFM, the most frequent CFSs are the same than for DBLP, but with slightly lower support rates. In this context, this means a large majority of users listened at least once to the same music than a friend (possibly a new one) not connected to other friends (or with different listening habits); and also stopped listening to the same music than a friend, or even ended the friendship relation. From the perspective of this system, this can be interpreted as the action of starting/stopping listening to music recommended by some other users.

In the case of Enron, births and deaths appear in the same time slice, followed by other births and deaths in a later time slice, for $99\%$ of the nodes. This means almost everyone started an email conversation with a new person or group twice during the considered period; and similarly, they stopped a conversation twice (not necessarily the same). When considering the event distribution (Figure~\ref{fig:eventNodeNumber}), we already observed births and deaths were the most common events. However, the CFSs show that not only some people, but almost everyone in all three networks adds or drops one (two for Enron) ego-component during the considered period.

\subsubsection{Longest Patterns}
The LFSs identified for each network are listed in Table~\ref{table:LongestPat}. In the case of Enron, the longest pattern spans $13$ time intervals (out of $45$). It is a sequence of births and deaths indicating the consecutive starts/ends of email conversations. There are a number of very similar shorter patterns alternating the same events (not represented in the table). They reflect the very unstable nature of the neighborhood in this dataset, as already noticed in the previous subsection. 

\begin{table}[!tb]
	\caption{Longest Sequential Patterns}
	\label{table:LongestPat}
	\centering
	\begin{tabular}{| l | l | r | r |}
		\hline
		Network & Sequential pattern $s$ & $|\mathcal{S}(s)|$ & $Sup(s)$ \\ 
        \hline
		DBLP & $\langle (B)(B)(E)(E)(E)(C)(C)(C)\rangle$  & 33 & 0.010  \\ 
        \hline
		LastFM &$\langle (B)(D, M, E)(D)(D,C)(C)(C)(C)\rangle$ & 14 & 0.008  \\ 
        \hline
        Enron &$\langle (B,D)(B)(D)(B)(D)(B)(D)(B)(D)(B)(B)(D)(B,D)\rangle$ & 1266 & 0.004  \\ 
        \hline
	\end{tabular}
\end{table}

\begin{figure*}[!tb]
 	\includegraphics[width=0.24\textwidth]{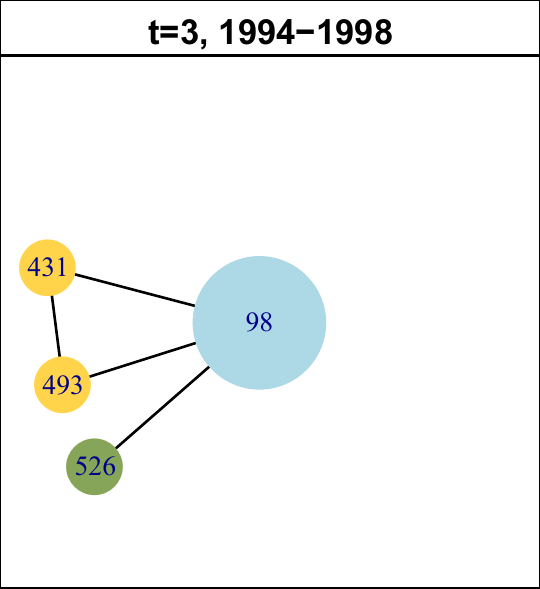}
 	\includegraphics[width=0.24\textwidth]{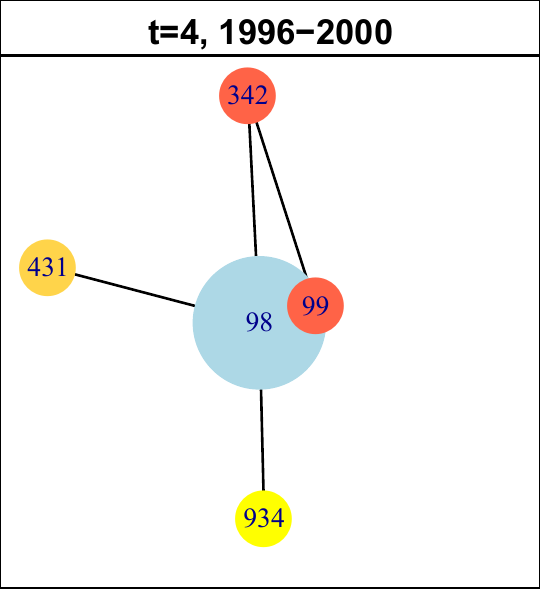}
 	\includegraphics[width=0.24\textwidth]{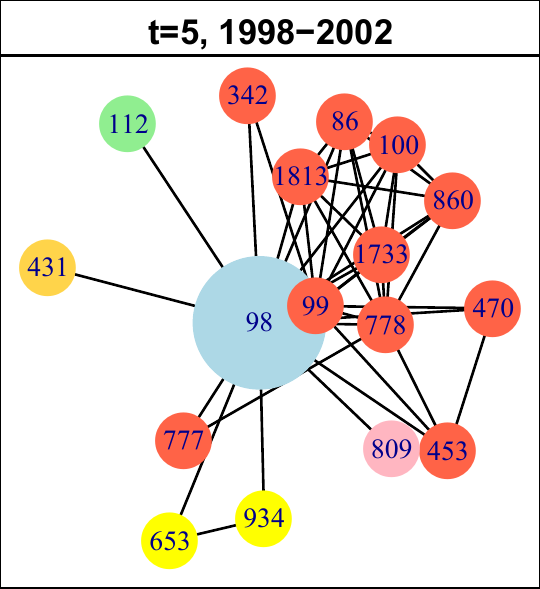}
 	\includegraphics[width=0.24\textwidth]{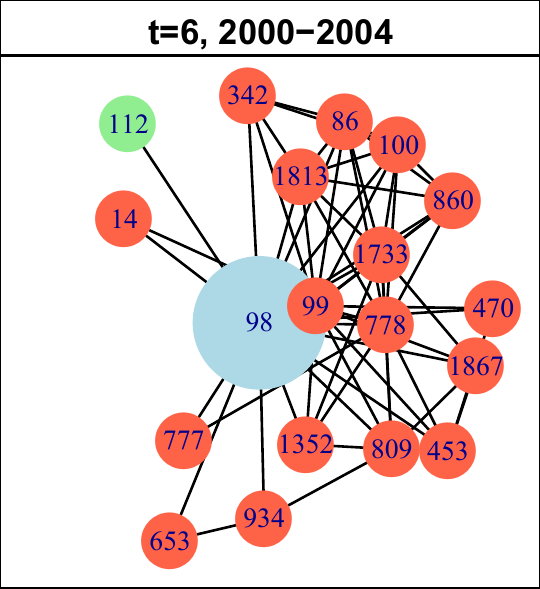}
    
 	\includegraphics[width=0.24\textwidth]{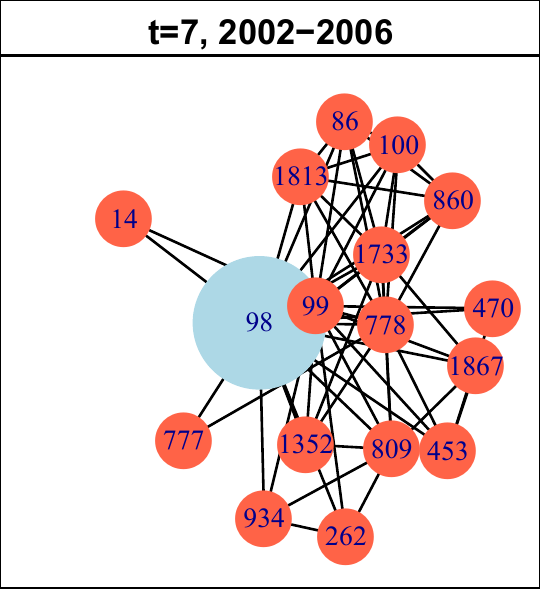}
 	\includegraphics[width=0.24\textwidth]{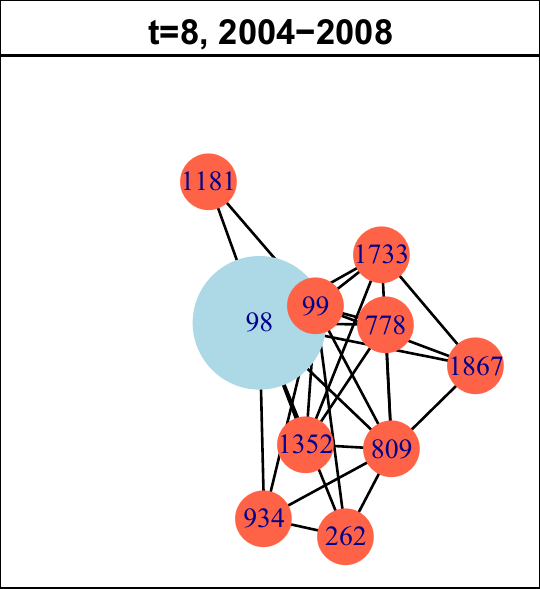}
 	\includegraphics[width=0.24\textwidth]{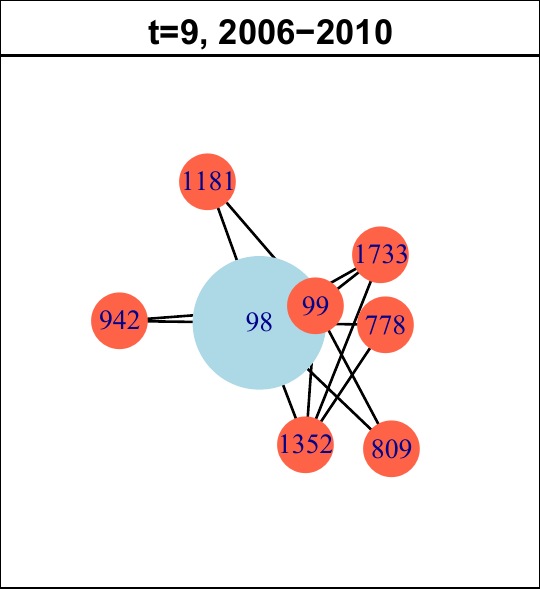}
 	\includegraphics[width=0.24\textwidth]{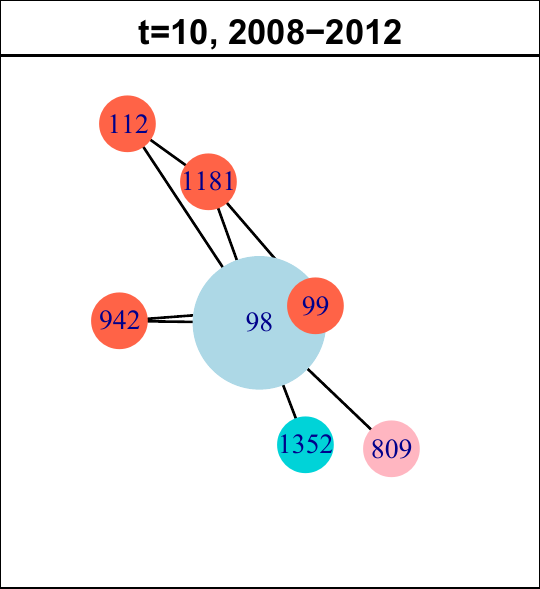}
 	\caption{Example of a node (in blue) following the longest frequent pattern in DBLP. Colors correspond to ego-components, and spatial positions are constant.}
 	\label{fig:longestPat}
\end{figure*}

In DBLP, the longest pattern spans $8$ time intervals (out of $9$). The trend is: first births (for $2$ time intervals), then expansions ($3$ times intervals) and finally contractions ($3$ time intervals). It concerns $1\%$ of the nodes, which is low, but we have found many very similar (though slightly shorter) patterns with higher support (not represented in the table). In total, all those nodes amount to approximately $11\%$ of the network. Hence, we can conclude that this is a significant trend in its evolution. It means authors tend to start new collaborations in the beginning of the considered period, then they develop their ego-components, before finally reducing their size. We do not observe significantly supported patterns containing merges or splits. The trend we identified is quite a common life cycle: creation, growth and decline. Usually, decline is followed by destruction (death) in many real-life phenomenons. However, in DBLP, it seems once a group of collaborators is set, it tends to last, even though it evolves. As an illustration of this observation, Figure~\ref{fig:longestPat} displays an example of such a life cycle for node $98$. For clarity, only the time slices relevant to the pattern are represented. The red ego-component initially containing nodes $99$ and $342$ is born between $t=3$ and $4$. Then, it expands ($t=5$ and $6$) and its size increases from $2$ to $17$ nodes. Later, it contractions ($t=8$ to $10$) back to $4$ nodes. This case is representative of many nodes in this network.

The longest pattern found in the LastFM network spans $7$ time intervals (out of $9$), and is supported by $14$ nodes. Unlike for DBLP, we did not detect any shorter pattern similar to this one, so we cannot affirm the trend concerns a significant number of nodes. This is because the long patterns are rare in LastFM, be them frequent or not. Put differently, there is not a lot of significant regularities in the way the nodes evolve (at least regarding their ego-components). The pattern shown in Table~\ref{table:LongestPat} is not unlike the one observed for DBLP: it starts with births, continues with expansions, itself followed by contractions. However, there are some differences: first, deaths are not limited to the end of the sequence, and second, merges appear together with expansions. After having investigated in detail the nodes following this trend, we noticed all of them have one big ego-component including many neighbors, and several small ego-components, many of them containing a single node. Expansions, merges and contractions affect the big ego-component, whereas births and deaths occur on the single neighbors. Interestingly, these $14$ nodes are the most active users on the LastFM service (for this dataset), as reflected by their degree: in average, it ranges from $42$ to $50$, while it is between $2.7$ and $3.3$ for the rest of the nodes. The patterns can be interpreted as follows: the users supporting it tend to listen to a relatively stable group of artists, together with their friends, which correspond to their core favorites. But they are also open to trying new artists, without necessarily sticking to them, as shown by the fluctuations observed in the pattern. This type of interpretation is characteristic of our event-based approach and could not be done through classic topological measures.

\begin{figure*}[!tb]
	\includegraphics[width=0.19\textwidth]{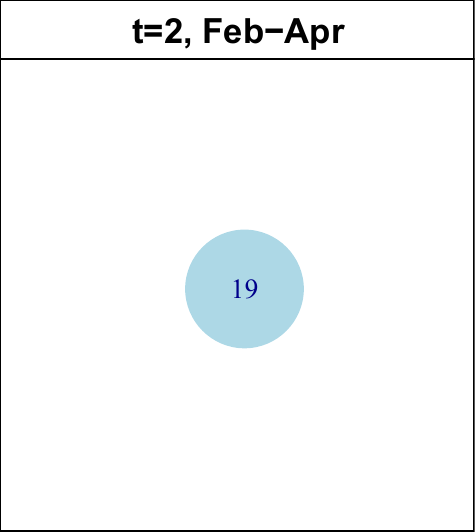}
	\includegraphics[width=0.19\textwidth]{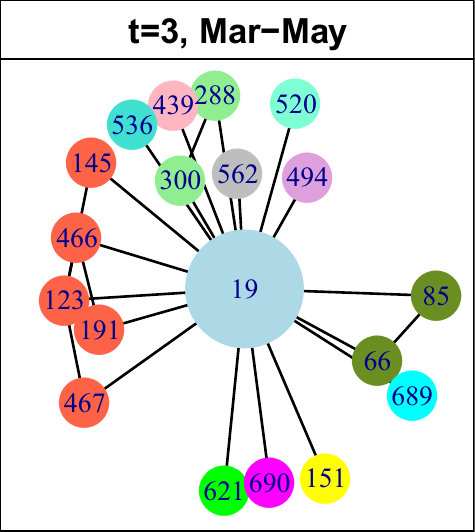}
	\includegraphics[width=0.19\textwidth]{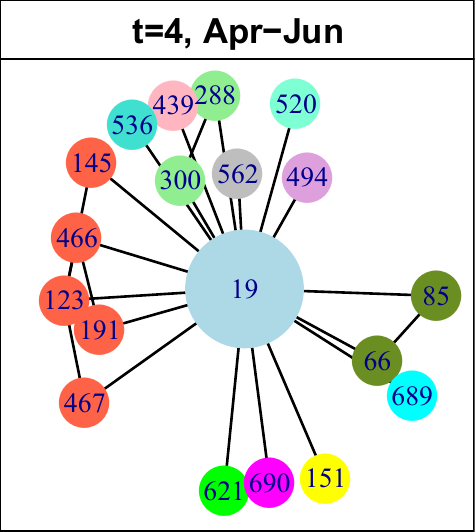}
	\includegraphics[width=0.19\textwidth]{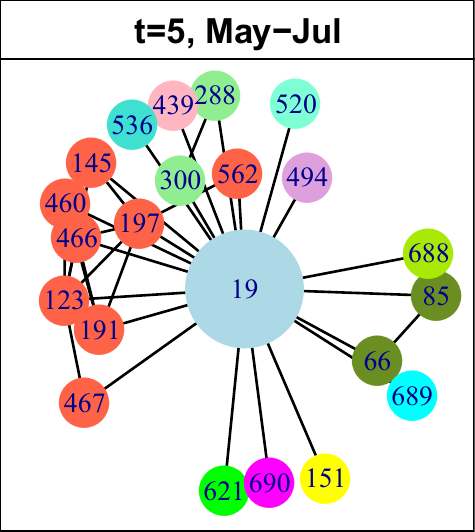}
	\includegraphics[width=0.19\textwidth]{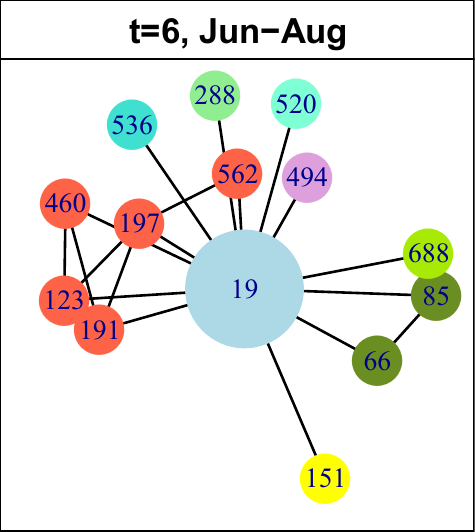}
    
	\includegraphics[width=0.19\textwidth]{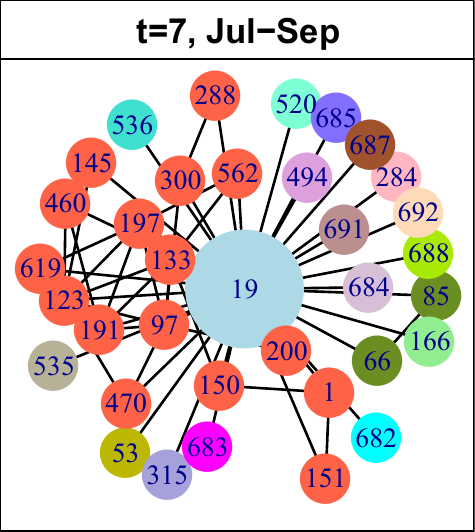}
	\includegraphics[width=0.19\textwidth]{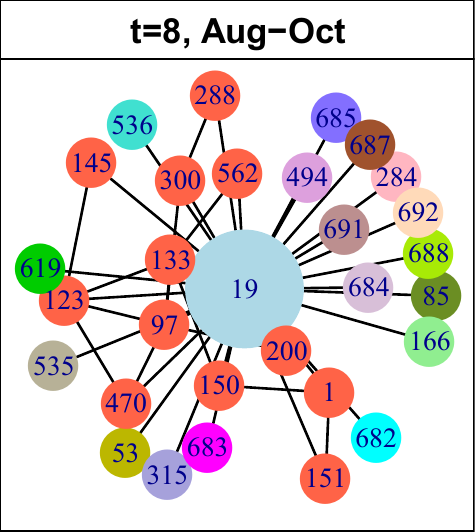}
	\includegraphics[width=0.19\textwidth]{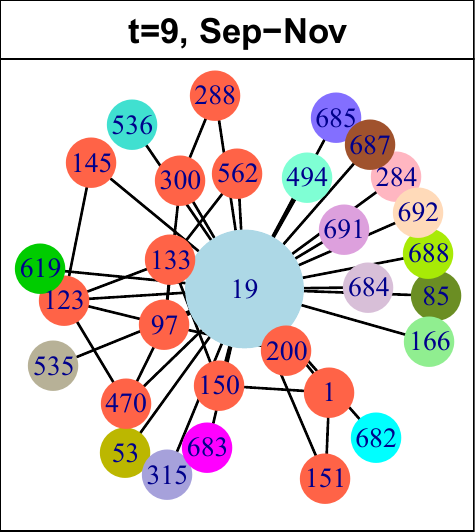}
	\includegraphics[width=0.19\textwidth]{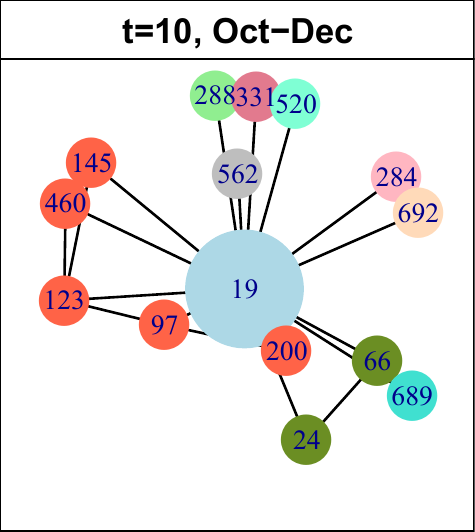}
	\makebox[0.19\textwidth][c]{}
 	\caption{Example of a node (in blue) following the most heterogeneous pattern in LastFM. Colors correspond to ego-components; spatial positions are constant.}
	\label{fig:variousPat}
\end{figure*}

Regarding Enron, the pattern we obtain spans $13$ time intervals and includes only births and deaths. In the context of this network, they mean joining/leaving an email conversation. But here, it is worth focusing on the events absent from the pattern, rather than on the present ones: there is no merge, split, expansion or contraction. For the Enron network, these events represent including/excluding a person to/from a conversation, or merging/separating different conversations. Since these do not appear in the pattern,this confirms that most of the users have conversations within stable groups. The profusion of deaths and births can be interpreted as the fact one user participates in several simultaneous conversations conducted at various time scales. 

The general conclusion of this subsection is that in these datasets, it was difficult to identify patterns being both long and largely supported. This can be due to a certain heterogeneity in the networks. To check this assumption, in the next section we deepen our analysis by mining patterns in subgroups of nodes obtained through a cluster analysis.

\subsection{Cluster Analysis}
\label{sec:ClusterAnalysis}
In this subsection, we present the results obtained through the cluster analysis described in Subsection \ref{sec:EventPatterns}, in particular we characterize them in terms of CFS and LFS, as we did previously for the whole network.

\subsubsection{Description of the Clusters}
\begin{figure*}[!tb]
	\center
 	\includegraphics[width=0.90\textwidth]{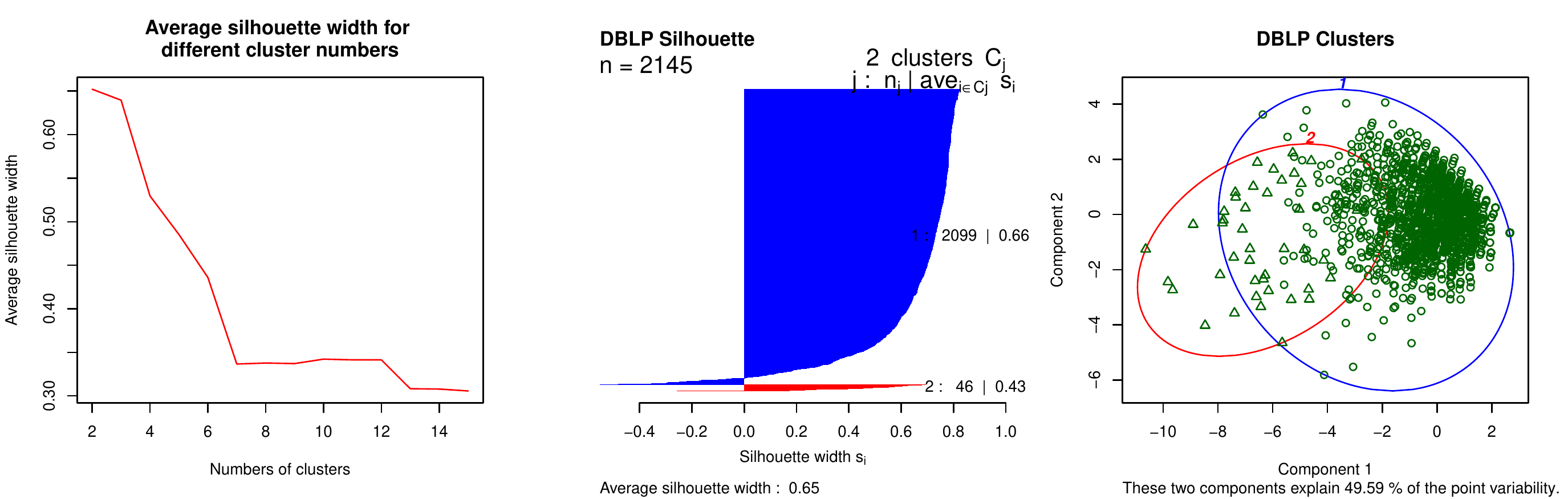} \\
 	\includegraphics[width=0.90\textwidth]{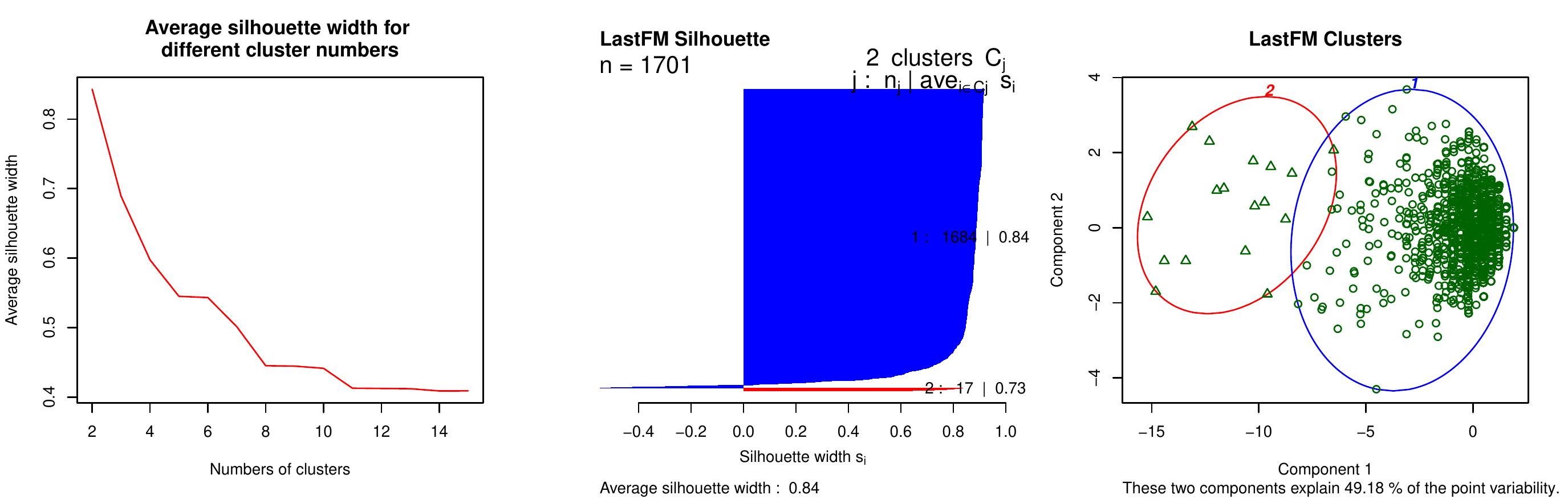} \\ 
 	\includegraphics[width=0.90\textwidth]{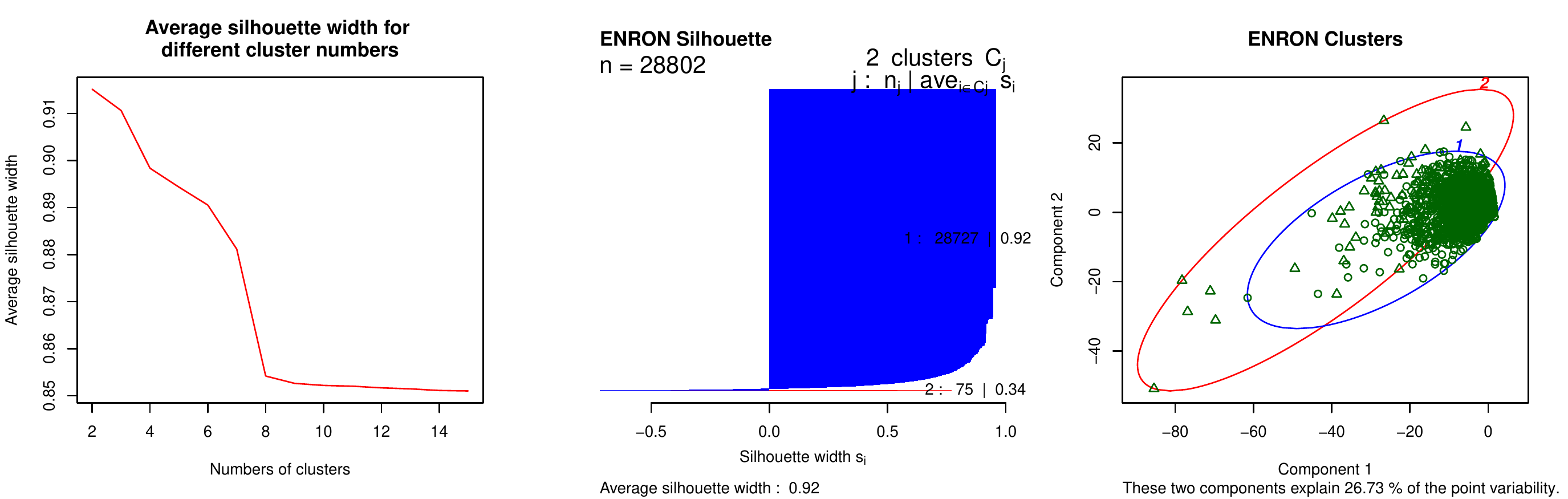}
 	
 	\caption{Cluster Results: average Silhouette width change (left) ; Silhouette width for two clusters (center) ; and Projection of the data points and clusters on two dimensions (right)}
 	\label{fig:Clusters}
\end{figure*}

As explained in Subsection \ref{sec:EventPatterns}, we apply a hierarchical clustering method to our set of sequential patterns. The left plots of Figure~\ref{fig:Clusters} show the average Silhouette width (ASW) as a function of the number of clusters, up to $15$ (after this value, the ASW stays low). The center plots show the Silhouette plot obtained for the best partition in terms of ASW, and the right plots display a projection of the data and clusters on a two-dimensional space. Table \ref{table:clusterDetails} shows some statistics describing the best partitions.

Interestingly, some results are common to all three considered networks. First, the best partitions contain $2$ clusters, and reach high ASW values (Table \ref{table:clusterDetails}), which means there is a clear separation in the data. Increasing the number of clusters decreases the ASW and causes the apparition of many small clusters, including numerous singletons. One of the two clusters is much larger than the other one, as shown in Table \ref{table:clusterDetails}. We call it Cluster 1 (in blue in Figure~\ref{fig:Clusters}), whereas the smaller one is Cluster 2 (in red). The most striking difference between them is the frequency and number of events their nodes undergo: in this sense, the nodes from Cluster 1 are much less active. As shown in Table \ref{table:clusterDetails}, only a very small proportion of Cluster 1 nodes undergo at least one event at each time slice of the considered period, whereas this observation is true for almost all nodes in Cluster 2. The number of nodes undergoing more than $10$ events over the complete time period is also much lower for Cluster 1.

\begin{table}[!tb]
	\caption{Statistics of Clusters}
	\label{table:clusterDetails}
	\centering
	\begin{tabular}{| l| l |l|l| l | l| l |}
    	\hline
		Network &Average    & cluster ID &\# of nodes &\# of nodes       &\# of nodes    & \# of nodes  \\ 
                &Silhouette &            &            &have event at all & with more     & with  negative\\ 
                &Width      &            &            &time interval     & than 10 event & Silhouette Width \\ 
        \hline
    	\multirow{2}{*}{DBLP} & \multirow{2}{*}{0.65}  & 1 & 2099  & 50  &  458 & 46 \\
	      &  & 2 & 46  & 38 &  46 & 2 \\
        \hline
                 
	    \multirow{2}{*}{LastFM} & \multirow{2}{*}{0.84}   & 1 &1684 &39  &141  &15  \\
	                            &                         & 2 &17 &16  &17  & - \\
          \hline                      
       \multirow{2}{*}{Enron} & \multirow{2}{*}{0.92}   & 1 &28727 & -  &4908  &160  \\
                               &                         & 2 &75 & -  &75  & 12 \\
		\hline
	\end{tabular}
\end{table}

Second, when considering the individual Silhouette widths of the nodes, we find that the majority of nodes in Cluster 1 reach a high value, meaning they belong with this cluster. Put differently, these clusters are very homogeneous (relatively to the criterion used for clustering). This appears quite clearly in the right plots of Figure~\ref{fig:Clusters} (blue clusters). Nevertheless, these clusters still contain a few nodes whose Silhouette width is negative, meaning they do not match their cluster well. Graphically, these nodes are located at the periphery of the clusters, near the center of the plots. The Clusters 2 are also very homogeneous, with only a very few nodes (or sometimes even none) with a negative Silhouette width. 

Based on these observations, we consider these clusters correspond to two classes of nodes characterized by different behaviors: \textit{stable nodes}, which are majority and whose neighborhood does not change much over time ; and \textit{active nodes}, different from most other nodes in their network, and who undergo numerous and frequent events. Nevertheless, as mentioned before, the clusters are not perfectly homogeneous, and some nodes are even located at the limit of the clusters in Figure~\ref{fig:Clusters}, indicating the possible existence of sub-clusters. Also, the level of separation between the clusters varies from one dataset to the other: they are clearly apart for LastFM, and much more intertwined for Enron. In order to better characterize our behavioral classes, we now look for their characteristic patterns.

\subsubsection{Patterns of the Clusters}
We obtain the patterns characterizing the behavioral classes by mining CFSs and LFSs in the corresponding clusters. For this purpose, we use the same methods as in Section~\ref{sec:sequential}, but with a restriction to the considered cluster. Table~\ref{table:clusterPatterns} lists some representative results. We get many similar patterns, with small differences, not only in terms of sequence, but also support $Sup(s)$ and growth rate $Gr(s)$. For the sake of simplicity, we only give the most representative ones in the table.

\begin{table}[!tb]
	\caption{Patterns found for the clusters}
	\label{table:clusterPatterns}
	\centering
	\begin{tabular}{ |l| l |l|l| l | l| l |}
    	\hline
		Network & Cluster & Pattern & $s$ & $|\mathcal{S}(s)|$ &$Sup(s)$ & $Gr(s)$  \\
                & ID      & Type    &     &                    &         &  \\ 
                
        \hline
    	\multirow{4}{*}{DBLP} & \multirow{2}{*}{1} & CFS & $\langle$(B)$\rangle$ &1995 & 0.95 & 0.95  \\
	     & & LFS  &$\langle$(B,D)(E)(E)(E)(E,C)(E)(C)$\rangle$ &32 & 0.01&0.07 \\ 
                 
       	 \cline{2-7}
	     & \multirow{2}{*}{2} & CFS& $\langle$(B,E)(E)$\rangle$  & 46 & 1.00 & 5.20 \\
	     & & LFS &$\langle $(E)(E)(E)(E)(E,C)(C)(E)(E)$\rangle$  & 18 & 0.39 & 821.35  \\
		
        \hline
		\multirow{4}{*}{LastFM} & \multirow{2}{*}{1} & CFS & $\langle$(D)$\rangle$ &1513 & 0.90 & 0.003  \\
	    &  & LFS  &$\langle$(B,D)(B)(D)(D)(B)$\rangle$ &67 & 0.04&0.048 \\

		\cline{2-7}
	    & \multirow{2}{*}{2} & CFS& $\langle$(S,C)(E)(S)(E)(E)$\rangle$  & 17 & 1.00 & 561.33 \\
	    & & LFS &$\langle $(B,D,S)(M,E)(B)(D)(D)(B)(B)$\rangle$  & 12 & 0.70 & $\infty$ \\

		\hline
        \multirow{4}{*}{Enron} & \multirow{2}{*}{1} & CFS & $\langle$(B,D)$\rangle$ &28574& 0.99  & 0.99  \\
	    &  & LFS  &$\langle$(B,D)(B)(D)(B)(B)(D)(B)(D)(D)$\rangle$ &2372 &0.082&0.086 \\ 
       
       	\cline{2-7}
	    & \multirow{2}{*}{2} & CFS& $\langle$(B,D)(D)(C)(B E)(B,E)(B,D,E,C)(B,D)$\rangle$ &75&1.00 &60.21  \\
	    & & LFS &$\langle$(B,D)(D)(D)(E,C)(B,E)(B)(B,D,C)(B,D)$\rangle$  &75 & 1.00 &55.99 \\
		\hline
	\end{tabular}
\end{table}

Like before, we observe certain similarities between the datasets. For the Clusters 1, the CFSs are exactly the same as the ones found for the whole graph (cf.~Table\ref{table:mostFreqPat}). Thus, despite their very high support, they are not characteristic of these clusters. This is confirmed by their growth rate, which is lower than $1$ and therefore indicates that the patterns are frequent not only in Cluster 1, but also in Cluster 2. The identified LFSs have, of course a much lower support. They are also less reminiscent of the ones found for the whole network (Table~\ref{table:LongestPat}), although they display a similar general form. Their growth rate is low, which means they are not specific to the Clusters 1 neither. But the interest of LFSs is to identify long trends, not necessarily highly supported ones. When focusing on the properties of the nodes supporting these long patterns, we identify two situations. On the one hand, in LastFM they have a very high Silhouette width, and therefore constitute the core of Cluster 1. When looking more thoroughly at these nodes, we can observe the events described by the pattern are generally caused by the creation/deletion of single links. On the other hand, in DBLP and Enron, their Silhouette width is very low, i.e. they are at the periphery of Cluster 1. Compared to the first situation, the events are caused by important changes in larger groups of nodes. Put differently, the second situation corresponds to nodes whose neighborhood is more unstable than for the rest of the cluster. This illustrates that LFSs can be used to characterize both the core or the periphery of such clusters.

Turning to the Clusters 2, we see much larger growth rates for all identified patterns, which means they are particularly typical of these clusters. The support is $1$ for all listed CFSs, meaning they are supported by all nodes in the concerned clusters. For DBLP, Cluster 2 is a small active group of nodes whose neighborhoods are increasing through expansions and births. For LastFM, the neighborhoods tend to expand too, but also to split. For Enron, the trend is long and include many births and deaths, as well as a few expansions and contractions. The LFSs also have a very large support, except for DBLP where it is less than half the nodes. For Enron, it is very similar to the CFS, so it does not bring much additional information. For LastFM, we get an infinite growth rate, which means the pattern is completely absent from Cluster 1. This pattern contains a variety of events, including split, merge and expansion. It is consistent with our observation from the previous subsection, in that it describes a behavior involving more neighborhood changes, compared to Cluster 1. For instance, the split reflects the fact that, for a user of interest, a group of friends separated in two groups with distinct music consumption habits, while keeping certain other habits similar to the user's.

Interestingly, for DBLP, the LFS is quite similar to that of Cluster 1, representing a long trend of expansion. Moreover, other nodes from Cluster 2 follow similar patterns, although not exactly the same. They all form a group of $24$ researchers often changing their coauthors. A detailed analysis reveals that most of them undergo at least $3$ events at each time slice, for an average total of $29$ events over the whole period, and that their ASW is $0.53$. In other terms, this group is active and representative of Cluster 2. It includes important senior researchers such as Jiawei Han, Rakesh Agrawal and Hongjun Lu. These are authorities of their domain, and all obtained their PhD degree before the 90s. The main qualitative difference between the LFSs of Cluster 1 and 2 is that the former starts with births and deaths. Further analysis shows most of the nodes supporting this Cluster 1 pattern undergo no event at all during the first time slice of the period, i.e. no neighborhood change in 1990-1996. They then undergo births and deaths during the next time slice (1992-1998) and expand their neighborhoods during the rest of the period. Some of the nodes supporting this pattern represent important researchers: Charu Aggarwal and Kaushik Chakrabarti. Their common point is they usually are very active but relatively young, all PhD students during the 90s. It seems these two groups of researchers follow similar patterns, but at different ages. Detecting such patterns could therefore help detecting promising researchers.

\section{Conclusion}
\label{sec:Conclusion}
In this work, we proposed a new method to characterize the evolution of dynamic networks at a local level. We defined the notion of neighborhood event, which corresponds to one among $6$ different changes in connectivity in the neighborhood of a node. We described a method to identify these events, and proposed to use sequential pattern mining to study them. We experimentally validated our tools on three real-world networks (DBLP, LastFM and Enron). First, we showed that counting the occurrences of the different types of events can give some insights on the global evolution of the network: all three networks are very different according to these descriptors, and we were able to catch historically significant periods in the case of DBLP and seasonal ones for Enron. Second, we used frequent pattern mining to identify certain trends among the nodes at the level of the network. The Enron trends reflect the routine of sporadically sending/receiving emails, whereas those of LastFM and DBLP describe a similar life cycle for ego-components: creation, growth and decline. Third, we performed a cluster analysis which allowed us to identify two types of behaviors appearing in all three networks: a very large cluster of nodes have a relatively stable neighborhood, whereas it undergoes much more diverse and frequent changes for a small cluster. We studied these clusters separately through pattern mining. It turns out the larger cluster is not homogeneous enough to be characterized, but parts of it can, such as core and periphery nodes. On the contrary, the smaller cluster is very homogeneous, and can even be described as a whole by LFSs possessing a very high support.

Our work can be seen as a first attempt to develop a tool allowing to characterize the evolution of dynamic networks at the local level of the node neighborhood. We believe it can be extended in various ways. From the theoretical perspective, we adopted a \textit{qualitative} approach to define and study node neighborhood evolution: first because we used various types of events, but also because we considered them in terms of presence/absence when mining the sequential patterns. However, as shown in our cluster analysis, it is also possible to consider node neighborhood evolution in a \textit{quantitative} way. So a first extension will be to design new topological nodal measures based on quantities such as the number of event occurrences by event type, the size of the ego-components undergoing them, and to adapt measures previously defined to describe community evolution \parencite{Toyoda2003,Palla2007}. Incidentally, the evolution of such measures would be computationally easier to analyze. A second extension will be to mine patterns defined not only in terms of community events, but including also other descriptors such as nodal topological measures (e.g. centralities) and attributes, as previously done to study communities in \parencite{Orman2015}.
The third extension concerns the evaluation of the significance of the detected events. This could be done by using a null model, which would allow comparing the number of events of a given type observed in the studied networks, with its expectation in the random model.
Fourth, our tool was designed for a descriptive task, but it could be used, either alone or in conjunction with some other descriptors, for inference. In particular, if the dynamic of the studied system can be characterized at the level of the node neighborhood, the event sequences could be used to perform link prediction. 

\section{Acknowledgment}
This article is supported by the Galatasaray University Research Fund (BAP) within the scope of the project number 14.401.002, and titled "Sosyal Ağlarda Küme Bulma ve Anlamlandırma: Zamana Bağlı Sıralı Örüntü Uygulaması".

\printbibliography

\end{document}